\documentclass[aps,prx,reprint,amsmath,amssymb,superscriptaddress,floatfix]{revtex4-2}
\usepackage[usenames]{color}
\usepackage{siunitx}
\usepackage{graphicx}
\usepackage[caption=false]{subfig}
\begin{document}

\title{Statistical modeling of adaptive neural networks explains coexistence of avalanches and oscillations in resting human brain}

\author{Fabrizio Lombardi}
\affiliation{Institute of Science and Technology Austria, Am Campus 1, A-3400 Klosterneuburg, Austria}
\author{Selver Pepi\'c}
\affiliation{Institute of Science and Technology Austria, Am Campus 1, A-3400 Klosterneuburg, Austria}
\author{Oren Shriki}
\affiliation{Department of Brain and Cognitive Sciences, Ben-Gurion University of the Negev, Beer-Sheva, Israel}
\author{Ga\v sper Tka\v cik}
\affiliation{Institute of Science and Technology Austria, Am Campus 1, A-3400 Klosterneuburg, Austria}
\author{Daniele De Martino}
\affiliation{Biofisika Institute (CSIC,UPV-EHU) and Ikerbasque Foundation, Bilbao 48013, Spain}


\begin{abstract}
Neurons in the brain are wired into adaptive networks that exhibit a range of collective dynamics. Oscillations, for example, are paradigmatic synchronous patterns of neural activity with a defined temporal scale.  Neuronal avalanches, in contrast, are scale-free cascades of neural activity, often considered as evidence of  brain tuning to criticality. While models have been developed to account for oscillations or avalanches separately, they typically do not explain both phenomena, are too complex to analyze analytically, or intractable to  infer from data rigorously. Here we propose a non-equilibrium feedback-driven Ising-like class of neural networks that simultaneously and quantitatively captures scale-free avalanches and scale-specific  oscillations. In the most simple yet fully microscopic model version we can analytically compute the phase diagram and make direct contact with human brain resting-state activity recordings via tractable inference of the model's two essential parameters. The inferred model quantitatively captures the dynamics over a broad range of scales, from single sensor oscillations and collective behaviors of nearly-synchronous extreme events on multiple sensors, to neuronal avalanches unfolding over multiple sensors across multiple time bins.  Importantly, the inferred  parameters correlate with model-independent signatures of ``closeness to criticality'', indicating  that the coexistence of scale-specific (neural oscillations) and scale-free (neuronal avalanches) dynamics in brain activity occurs close to a non-equilibrium critical point at the onset of self-sustained oscillations.
\end{abstract}  

  \maketitle

 \section{Introduction}
   
Synchronization is a key organizing principle that leads  to the emergence of  coherent macroscopic behaviors across diverse biological networks~\cite{pikovsky_sync}. 
From  Hebb's ``neural assemblies''~\cite{pla:hebb} to synfire chains~\cite{abeles_1982,corticonics},  synchronization has also strongly shaped our understanding of brain dynamics and function~\cite{buz_syntax}. The classic and arguably most prominent example of large scale neural synchronization are brain oscillations, first reported about a century ago~\cite{berg:alpha}: periodic, large deflections in electrophysiological recordings, such as electroencephalography (EEG), magnetoencephalography (MEG), or local field potential (LFP)~\cite{berg:alpha,buz:osc}. Because oscillations are thought to play a fundamental role in brain function, their mechanistic origins have been the subject of intense research.  According to the current view, the canonical circuit that generates prominent brain rhythms such as the alpha oscillations and the alternation of up- and down-states utilizes mutual coupling between excitatory (E) and inhibitory (I) neurons~\cite{wilsoncowan1972,wang_2010_osc,freyer_2012_jneuro}. Alternative circuits, including I-I population coupling, have been proposed to explain other brain rhythms such as high-frequency gamma oscillations~\cite{chow1998sync,borgers2005ei,buz_2012_gamma}. Setting biological details aside, the majority of research has predominantly focused on the emergence of synchronization at a preferred temporal scale --- the oscillation frequency.

Brain activity also exhibits complex, large-scale cooperative dynamics with characteristics that are antithetic to those of oscillations. In particular, empirical observations of ``neuronal avalanches''  have shown that brain rhythms coexist with activity cascades in which neuronal groups fire in patterns with no characteristic time or spatial scale, suggesting that the brain may operate near criticality~\cite{plenz:pl03,iit:ava,plenz:awa,plenz:tbg,taglia_2012_point,shriki13,priesemann2013sleep,fl2012,frontiers,fl2014_epjst,fontanele_2019_prl,poncealvarez_2018}. In this context, the coexistence of scale-free neuronal avalanches with scale-specific oscillations  suggests an intriguing dichotomy that is currently not understood. On the one hand,  models of brain oscillations are very specific and seek to capture physiological mechanisms underlying particular brain rhythms. On the other hand, attempts to explain the emergence of neuronal avalanches almost exclusively focus on  criticality-related aspects and ignore the coexisting behaviors such as oscillations, even though they themselves may be constitutive for understanding the putative criticality. Among the few exceptions ~\cite{poil_2012_jneuro,silvia_2013_bio,disanto_2018_pnas,kinouchi2017entropy,kinouchi2019scirep},  Poil et al  proposed a probabilistic integrate and fire (IF) spiking model with E and I neurons that  generates long-range correlated fluctuations reminiscent of MEG oscillations in the resting state, with supra-threshold activity following power-law statistics consistent with neuronal avalanches and criticality~\cite{poil_2012_jneuro}. More recently, by adopting  a coarse-grained Landau-Ginzburg approach to neural network dynamics, Di Santo et al have shown that neuronal avalanches  and related putative signatures of criticality co-occur at a synchronization phase transition, where collective oscillations may also emerge~\cite{disanto_2018_pnas}. These results were successively extended to a hybrid-type synchronization transition in a generalized Kuramoto model~\cite{buendia_2021_prr}. 

While both these and other proposed approaches show that neuronal avalanches may coexist with some form of network oscillations \cite{poil_2012_jneuro,kinouchi2019scirep} or network synchronization \cite{disanto_2018_pnas,buendia_2021_prr},   
they suffer from  three major shortcomings. First, these models are  neither simple (e.g., in terms of parameters) nor analytically tractable, making an exhaustive exploration of their phase diagram out of reach. Second, none of the two models simultaneously captures events at the microscopic scale (individual spikes) and macroscopic scale (collective variables). Third, it is not clear how to connect these models to data rigorously, beyond relying on qualitative correspondences.

Here we propose a minimal, microscopic, and analytically tractable  model class that can capture a wide spectrum of emergent phenomena in brain dynamics, including neural oscillations, extreme event statistics,  and scale-free neuronal avalanches~\cite{plenz:pl03}.
\textcolor{black}{This model class is inspired by the recent theoretical observation that systems with many interacting degrees of freedom and   coexistence of distinct phases may  develop self-oscillations in the presence of feedback loops between  control and order parameters~\cite{ddemartino_jphys_2019}. Brain dynamics, with its patterns of state transitions~\cite{lo2004common,freyer_2009}, long-range correlated fluctuations and cooperative behaviors showing marks of criticality~\cite{linken01,plenz:awa,fl2020_lrtc,fl2020jneurosci,tkavcik2015thermodynamics},  relies on a number of regulatory feedback loops that are considered at the core of its function~ \cite{freyer_2012_jneuro,deco_2017_scirep}. Here, we hypothesize that basic feedback mechanisms controlling the excitability of the system   could produce the coexistence of the antithetic scale-specific  oscillations and scale-free avalanches in brain activity.   
To test this hypothesis, we build upon the well-established   analogy between neural networks and spin systems~\cite{cragg1954ising,hopfield1982neural,schneidman2006weak,roudi2009pre,tkacik2010optimal,tkavcik2015thermodynamics}, and put forward a non-equilibrium extensions of the Ising model of statistical physics with an extra feedback loop which enables self-adaptation. As a consequence of feedback, neuronal dynamics is driven by the ongoing network activity, generating a rich repertoire of dynamical behaviors.} \textcolor{black}{As in previous applications of the static Ising model to neural systems \cite{schneidman2006weak, roudi2009pre,tkacik2010optimal}, spins model the spiking, binary nature of individual neurons (firing versus silent). Thus, the structure of the simplest model from this class permits microscopic network dynamics investigations as well as analytical mean-field solution in the Laudau-Ginzburg spirit, and in particular, allows us to construct the model's phase diagram.}

The tractability of our model enables us to make direct contact with MEG data on the resting state activity of the human brain. With its two free parameters inferred from data, the model closely captures brain dynamics across scales, from single sensor MEG signals  to collective behavior of extreme events and neuronal avalanches. Remarkably, the inferred  parameters 
indicate that scale-specific (neural oscillations) and scale-free (neuronal avalanches) dynamics in brain activity coexist close to a non-equilibrium critical point that we proceed to characterize in detail. 

\section{Results}

\subsection{Adaptive Ising model}
\begin{figure}
\centering
\includegraphics[width=1\linewidth]{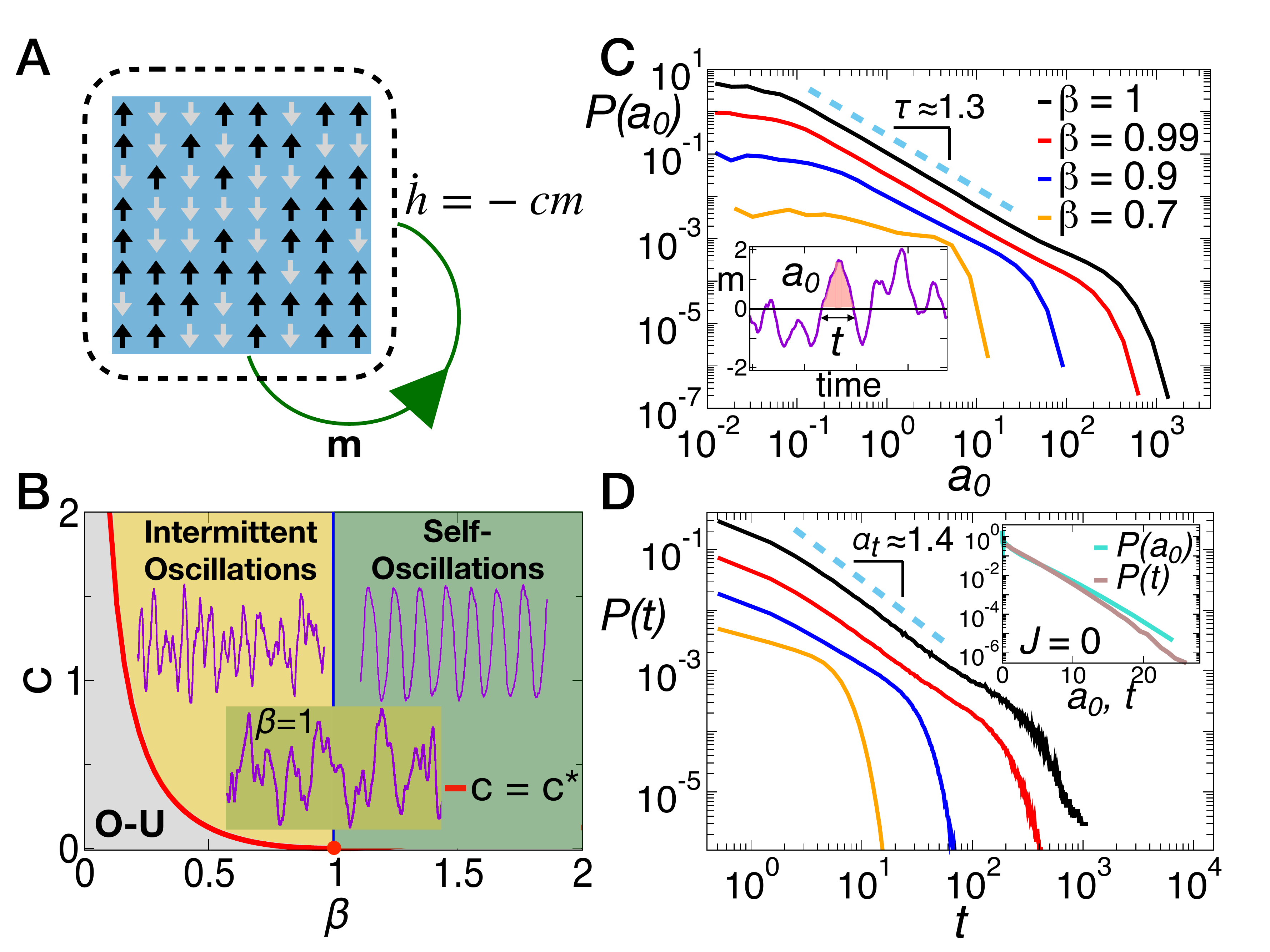}
\caption{Adaptive Ising model exhibits coexistance of oscillations and scale-free activity excursions near the critical point. (A) Schematic illustration of the model.  Interacting  spins $s_i$ ($i = 1,2,...,N$) take values $+1$ (up arrows) or $-1$ (down arrows) and experience a time-varying external field $h(t)$ that mimics an activity-dependent feedback mechanism. (B) Phase diagram for the mean-field adaptive Ising model.  An Andronov-Hopf bifurcation at $\beta_c=1$ separates self-sustained oscillations in the total activity $m(t)$  for $\beta> \beta_c$ (green) from the regime of intermittent oscillations  (yellow) for $c$ above $c^*(\beta)$ (solid red line) and an Ornstein-Uhlenbeck process for $c$ below $c^*$  (gray).  (C) Reversal time $t$ is the time interval between consecutive  zero-crossing events in $m$ and $a_0$ is the area under the $m(t)$ curve between two zero-crossing events (inset).  Distributions $P(a_0)$ are shown in the resonant regime, $c > c^*$, for different values of $\beta$. When $\beta \approx 1$, $P(a_0)$ is approximately power-law with exponent $\tau = 1.29 \pm 0.01$. (D) Distributions $P(t)$ of the reversal times are shown in the resonant regime, $c > c^*$, for different values of $\beta$. When $\beta \approx 1$, $P(t)$ is approximately power-law with exponent $\alpha _t = 1.40 \pm 0.01$. Inset: Distributions $P(a_0)$ and $P(t)$ for  the uncoupled model, $J = 0$, always exhibit exponential instead of  power-law behavior (note linear horizontal scale). }
\label{fig:1}
\end{figure}

We consider a population of interacting neurons whose dynamics is self-regulated by a time-varying field that depends on the ongoing population activity level~(Fig.~\ref{fig:1}A). The $N$ spins $s_i = \pm 1$ ($i = 1,2,...,N$, $N=10^4$ in our simulations unless specified differently) represent excitatory neurons that are active when $s_i = + 1$ or inactive when $s_i = -1$. In the simplest, fully  homogeneous scenario described here, neurons interact with each other through  synapses of equal strength $J_{ij} = J = 1$. 
%
%
 The ongoing network  activity is defined as $m(t) = \frac{1}{N}\sum_{i=1} ^N s_i (t)$ (i.e., as the magnetization of the Ising model) and each neuron experiences a uniform negative feedback $h$ that depends on the network activity as $\dot{h} = -cm$, with $c$ determining the strength of the feedback.
Neurons $s_i$ are stochastically activated according to the Glauber dynamics, where the new state of neuron $s_i$ is drawn from the marginal Boltzmann-Gibbs distribution 
$P(s_i) \propto \exp(\beta \tilde{h}_is_i)$, with $\tilde{h}_i = \sum_{j\neq i} J_{ij} s_j + h$, where $\beta$ is reminiscent of the inverse temperature for an Ising model (see Appendix B).

Multiple interpretations of this model are possible. On the one hand, negative feedback can be identified with a mean-field approximation to the inhibitory neuron population that uniformly affects all excitatory neurons with a delay given by the characteristic time $c^{-1}$ (see Appendix B). On the other hand, feedback could be seen as intrinsic to excitatory neurons, mimicking, e.g., spike-threshold adaptation~\cite{azouz:99,azouz:2000,henze2001_spikeths,wilent2005_spikeths}. Exploration-worthy (and possibly more realistic) extensions within the same model class are accessible by considering two ways in which geometry can enter the model. First, as in the standard Ising magnet, the interactions $J$ can be restricted to simulate local excitatory connectivity, e.g., to nearest neighbors on a 2D lattice. Second, feedback $h_i$ to neuron $i$ could be derived from a local magnetization in a neighborhood around neuron $i$ instead of the global magnetization; in the interesting limiting case where $\dot{h_i} = -c s_i$,  each neuron would feed back on its own past spiking history only, and the model would reduce to a set of coupled ``binary oscillators''~\cite{ddemartino_jphys_2019}. Irrespective of the exact setting, the model's mathematical attractiveness stems from its tractable interpolation between stochastic (spiking of excitatory units) and  deterministic (feedback) elements.

Network behavior is determined by feedback strength $c$ and inverse temperature $\beta$. In the fully-connected continuous-time limit, the model can be described with the following Langevin equations:
\begin{eqnarray}\label{eq:sde}
\dot{m} &=&  -m +\tanh\left[\beta(Jm+h)\right] + b \xi \\
\dot{h} &=& -c m,  \nonumber
\end{eqnarray}
where $\xi$ is unit uncorrelated Gaussian noise; the stochastic term thus has amplitude $b =  \sqrt{2/(\beta N)}$.
Equations~(\ref{eq:sde}) can be linearized around the stationary point $(m^*=0,h^*=0)$ to calculate dynamical eigenvalues and construct a phase diagram (Fig.~\ref{fig:1}B):
\begin{eqnarray}
\lambda_{\pm} &=& \frac{(\beta-1)}{2} \pm \frac{\sqrt{(\beta - 1)^2 - 4c\beta}}{2}.
\label{eq:eigenvalue}
\end{eqnarray} 
For $c = 0$, $h = 0$, the model reduces to the standard infinite-dimensional (mean field) Ising model with a second order phase transition at $\beta = \beta_c=1$. At non-zero feedback, $c>0$, the model is driven out of equilibrium and its critical point at $\beta_c$ coincides with an Andronov-Hopf bifurcation~\cite{izhikevich2007dynamical,ddemartino_jphys_2019}. For $c$ below a threshold value $c^*=(\beta -1)^2/4\beta$, $m(t)$ is described by an Ornstein-Uhlenbeck process (O-U) independently of $\beta$. For $\beta < \beta_c$, the system is stable and shows a crossover from a stable node with exponential relaxation (two negative real eigenvalues) to a stable focus with oscillation-modulated exponential relaxation (two imaginary eigenvalues; ``resonant regime'') when $c$ increases beyond $c^*$~(Fig.~S1). In the resonant regime, $c > c^*$,  oscillations become more prominent as the critical point $\beta_c = 1$ is approached, finally transitioning into self-sustained  oscillations for $\beta > \beta_c$~(Fig.~S2).

We focus on the resonant regime below and at the critical point, and study the reversal times and zero-crossing areas of the total network activity $m(t)$~(Fig.~\ref{fig:1}C). The distribution $P(a_0)$ of the zero-crossing area follows a power-law behavior with an exponent $\tau = 1.29 \pm 0.01$ in the vicinity  of the critical point. As $\beta$ decreases,  the scaling regime shrinks until it eventually vanishes for small enough $\beta$. Similar behavior is observed for the distribution $P(t)$ of reversal times. This distribution also follows a power-law with an exponent $\alpha _t = 1.40 \pm 0.01$ near the critical point~(Fig.~\ref{fig:1}D). Both distributions have an exponential cutoff related to the characteristic time of the network activity oscillations, $1/c$; this cutoff transforms into a hump as $\beta \rightarrow 1$ and  $c \gg c^*(\beta)$, i.e., as oscillations in $m(t)$ become increasingly prominent~(Fig.~S3).
Importantly, for the non-interacting ($J = 0$) model, the distributions $P(a_0)$ and $P(t)$ follow a purely exponential behavior~(Fig.~\ref{fig:1}D, inset), indicating that the coexistence of oscillatory bursts and power-law distributions for the network activity requires neuron interactions as well as the adaptive feedback~(Fig.~S4).

\subsection{Model inference from local resting-state brain dynamics}  
In the resonant regime below the critical point  ($c > c^*,\beta<\beta_c$), it is possible to analytically compute the autocorrelation function of the ongoing network activity $m(t)$ in the linear approximation~\cite{gardiner:stoch}:
\begin{equation}
C (\tau) = e^{-\gamma \tau}(\cos\omega \tau + \frac{\gamma}{\omega}\sin\omega \tau), 
\label{eq:autocorr}    
\end{equation}
where $\gamma = (1-\beta)/2$ and $\omega = \sqrt{\beta c - (1- \beta)^2/4}$. The autocorrelation $C(\tau)$ can be used to infer model parameters $\beta$ and $c$ from empirical data by moment matching, thereby locating the observed system in the phase diagram~(Fig.~\ref{fig:1}B). 

\textcolor{black}{We test the proposed approach on MEG recordings of the awake resting-state of the human brain (Appendix A). We first analyze brain activity on individual MEG sensors. To this end, we compare the   magnetic field recorded on individual MEG sensors with the magnetization $m$ of the model~(Fig.~\ref{fig:1}). This analogy relies on the nature of the brain magnetic fields captured by the MEG, which are generated by synchronous postsynaptic currents in cortical neurons \cite{meg2012review}. 
Because the intensity of such currents depends on the number of actively firing neurons, the temporal fluctuations in the MEG signal are related to average firing rate. Similarly, in the model, the fluctuations in the magnetization $m$ are directly related to the number of active neurons over time, that is the firing rate.}

\begin{figure*}
\centering
\includegraphics[width=1\linewidth]{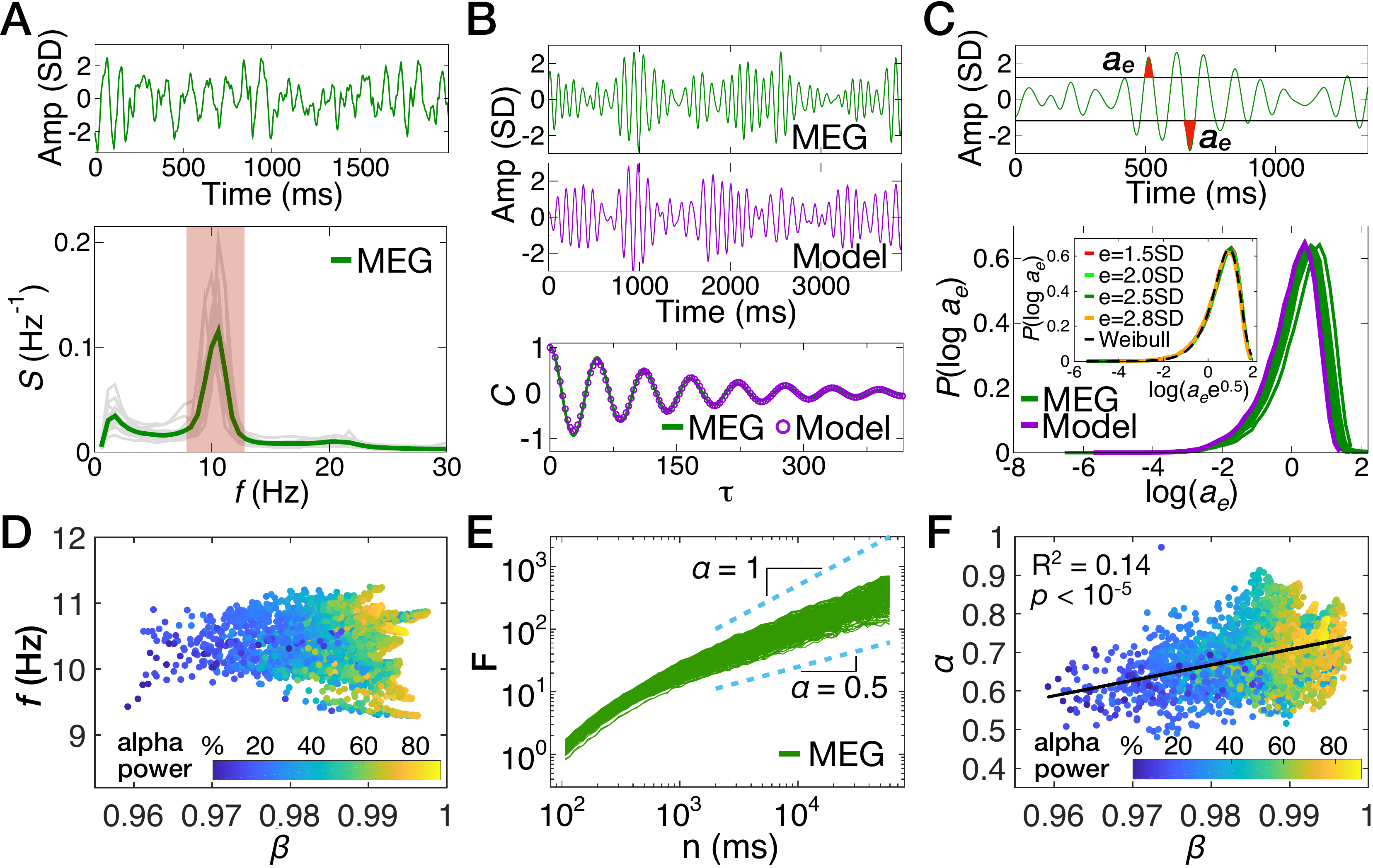}
\caption{MEG resting state activity of the human brain corresponds to a marginally sub-critical adaptive Ising model. (A) Example trace from a single MEG sensor (top) predominantly contains power in the alpha band (8-13 Hz; bottom, shaded region). Power spectra of MEG signals (bottom; gray = average across 273 MEG sensors for each of the 14 subjects; green = average over sensors and subjects) peak around 10 Hz. (B) Example alpha bandpass filtered MEG signal (top; green) and the simulated total activity $m(t)$ of a model with parameters matched to data (top; violet) show qualitatively very similar behavior. Model parameters ($\beta = 0.9870$ and $c = 0.1129$ for this trace) are inferred by fitting the analytical form of the autocorrelation function $C(\tau)$~(bottom; green line), to autocorrelation estimated from MEG data (bottom; violet dots).   (C) Schematic of the area under the curve $a_e$ (red) for a given threshold $\pm e$ in units of signal SD (top). Distributions $P(\log a_e)$ of the logarithm of the area under the curve $a_e$, with $e = 2.5$SD, for MEG data (green curves = average over sensors for each subject) and the model (violet curve = simulation at baseline parameters, see text). Inset: Rescaled distributions of $a_e$ collapse to a universal Weibull-like distribution across different threshold values $e$ (Weibull parameters: $k = 1.74$, $\lambda = 2.58$).  (D) Central frequency $f = \omega/2\pi = (\beta c - (1- \beta)^2)^{1/2}/8\pi$ of the fitted model plotted against fitted $\beta$,  across all MEG sensors and subjects (color = fraction of total MEG signal power in the alpha band). $\beta$ values closer to critical $\beta_c=1$ are correlated with higher power in the alpha band ($R^2 = 0.59$; $p < 0.001$).  (E) Root-mean-square fluctuation function $F(n)$ of the DFA for the amplitude envelope of MEG sensor signals in the alpha band (green lines = individual sensors for a single subject). $F(n)$ scales as  $F(n) \propto n^{\alpha}$ for $2$ s $<n<60$ s (dashed lines), with $\alpha > 0.5$ for all MEG sensors ($0.53 < \alpha < 0.85$).  (F) Inferred $\beta$ values correlate with the corresponding DFA exponents $\alpha$ for all MEG sensors and subjects. } 
\label{fig:2}
\end{figure*}

During resting wakefulness the brain activity is largely dominated by oscillations in the alpha band ($8-13$ Hz) (Fig.~\ref{fig:2}A), which has been the starting point of many investigations~\cite{buz:osc,freyer_2009,linken12,clayton_2017_alpha} including ours reported below; similar results are also obtained for the broadband activity (Fig.~S5).
After isolating the alpha band, we estimate the quantities $\gamma$ and $\omega$ by fitting the empirical $C(\tau)$ to the functional form given by Eq~(\ref{eq:autocorr}). Fig.~\ref{fig:2}B illustrates the typical quality of the fit and the qualitative resemblance between the model and MEG sensor signal dynamics.

Since our model is fit to reproduce the second-order statistical structure in the signal, we next turn our attention to signal excursions over threshold, a higher-order statistical feature routinely used to characterize bursting brain dynamics~\cite{freyer_2009,taglia_2012_point,fl2013,palva13,wang2019plos,fl2020jneurosci}.
To that end, we construct the distribution of (log) areas under the signal above a threshold $\pm e$~(Fig.~\ref{fig:2}C)~\cite{freyer_2009}. $P(\log a_e)$ is bell-shaped, featuring strongly asymmetric tails for  MEG sensors as well as the model (Fig.~\ref{fig:2}C).  Variability across subjects is mostly related to signal amplitude modulation, resulting in small horizontal shifts in $P(\log a_e)$ but no variability in the distribution shape. Remarkably, the rescaled distribution is independent of the threshold $e$ over a robust range of values, and is well-described by a Weibull form, $P_W(x;\lambda,k) = \frac{k}{\lambda}(\frac{x}{\lambda})^{k-1} e^{-(x/\lambda)^k}$~(Fig.~\ref{fig:2}C, bottom panel inset; Fig.~S6). Taken together, these observations indicate that our model has the ability to capture non-trivial aspects of amplitude statistics in MEG signals, within and across different subjects (Fig.~S7).

Parameters inferred across all sensors and subjects suggest  baseline values of $\beta = 0.99$ and $c = 0.01$ that are well matched to data, which we use for all subsequent analyses (unless stated otherwise). Specifically, we find the best-fit $\beta$ values strongly concentrate in a  narrow range around $\beta \approx 0.99$ ($\beta = 0.986 \pm 0.006$; $c = 0.012 \pm 0.001$), very close to the critical point (Fig.~\ref{fig:2}D and Fig.~S8). Even though all analyzed signals are bandpass-limited to a central frequency around 10 Hz by filtering, closeness to criticality appears to strongly correlate with the fraction of total power in the raw signal in the alpha band ($R^2 = 0.59$; $p < 0.001$). This suggests that alpha oscillations may be closely related to critical brain tuning during the resting state~\cite{linken01,linken12,palva13,shriki13,fl2020_lrtc}. 

A classic fingerprint of tuning to criticality is the emergence of long-range temporal correlations (LRTC), which have been documented empirically~\cite{linken01,linken12,palva13,berthouze10,meisel2017lrtc,fl2020_lrtc,fl2020jneurosci}. LRTC in the alpha band have been investigated primarily by applying the detrended fluctuations analysis (DFA) to the amplitude envelope of MEG or EEG signals in the alpha band (Appendix C)~\cite{linken01,poil_2012_jneuro,palva13, ros_nfb_2014}. Briefly, DFA estimates the scaling exponent $\alpha$ of the root mean square fluctuation function $F$ in non-stationary signals with polynomial trends~\cite{peng:1994}. For signals exhibiting positive (or  negative) LRTC, $F$ scales as $F \propto n^\alpha$ with $0.5 <\alpha < 1$ (or $ 0 < \alpha < 0.5$, respectively); $\alpha = 0.5$ indicates the absence of long range correlations;  $\alpha$ also approaches unity for a number of known model systems as they are tuned to criticality \cite{ellis1978curie,kertesz2008fluct}.

To test for the presence of LRTC using DFA, we analyzed the scaling behavior of fluctuations and extracted their scaling exponent $\alpha$~\cite{linken01,linken12}. To avoid spurious correlations introduced by signal filtering, $\alpha$ was estimated over the range $2\;\mathrm{s} < n < 60\;\mathrm{s}$~(Fig.~\ref{fig:2}E)~\cite{linken01,linken12}. We find that $\alpha$ is consistently between 0.5 and 1 for all MEG sensors and subjects, in agreement with previous analyses~\cite{linken01,montez09,palva13,palva2015jn,meisel2017lrtc,berthouze10}.
Importantly, model-free $\alpha$ values measured across MEG sensors positively correlate with the inferred $\beta$ values from the model~(Fig.~\ref{fig:2}F), indicating that higher $\beta$ values are diagnostic about the presence of long-range temporal correlations in the amplitude envelope. 

Taken together, our analyses so far show that the adaptive Ising model recapitulates single MEG sensor dynamics by matching their autocorrelation function and the distribution of amplitude fluctuations, and further suggest that the true MEG signals are best reproduced when the adaptive Ising model is tuned close to, but slightly below, its critical point ($\beta \lesssim 1$).

\subsection{Scale invariant collective dynamics of extreme events}

We now turn our attention to phenomena that are intrinsically collective: \emph{(i)} coordinated supra-threshold bursts of activity, which emerge jointly with LRTC in alpha oscillations~\cite{poil_2012_jneuro,palva13}; and \emph{(ii)} neuronal avalanches, i.e., spatio-temporal cascades of threshold-crossing sensor activity, which have been identified in the MEG of the resting state of the human brain~\cite{shriki13,fl2020_lrtc}. Both of these phenomena are generally seen as chains of extreme events that are diagnostic about the underlying brain dynamics~\cite{fraiman12,taglia_2012_point,fl2013,oshrit2016jn,oren18,fl_book_19}.

\begin{figure*}
\centering
\includegraphics[width=1\linewidth]{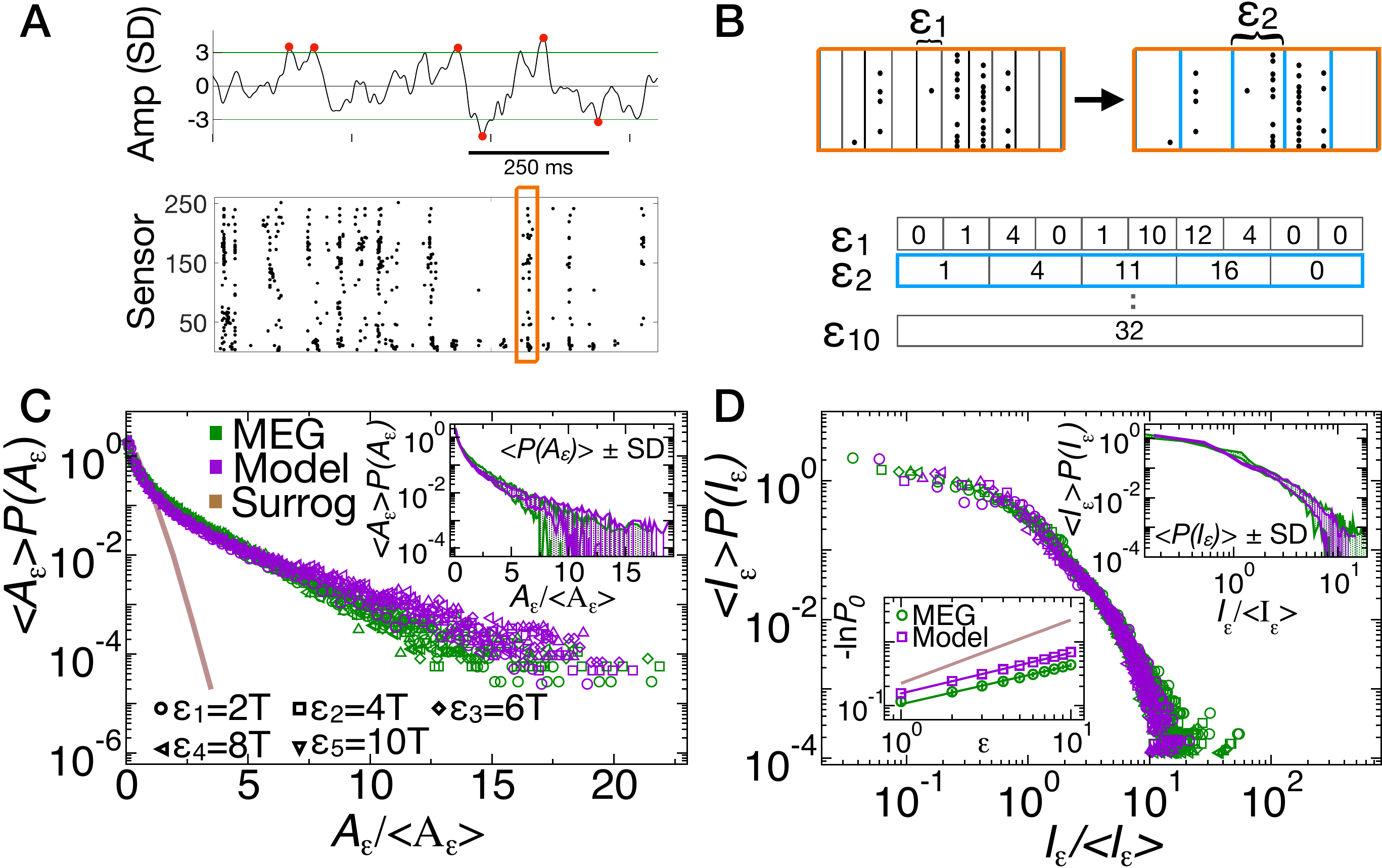}
\caption{Non-exponential extreme event statistics in MEG resting state activity are reproduced by a marginally subcritical adaptive Ising model. (A) Extreme events on a single sensor are defined as extreme signal excursions (top; red dots) crossing a threshold $e = \pm n$SD (horizontal lines). Resulting raster of extreme events shown across $273$ MEG sensors of a single subject across approximately 500 ms  recording (bottom). (B) Events are grouped together in temporal bins $\epsilon_n=nT$ in multiples of the sampling interval $T$ (top), to define instantaneous network excitation $A_{\epsilon}$, the total number of extreme events across all sensors in a time bin. Representative sequences of network excitation extracted from the raster in the top panel for increasing bin size  $\epsilon_n$ (bottom). (C) Rescaled distribution of network excitation, $P(A_\epsilon)$,  for $e = 2.9$SD and a range of bin sizes $\epsilon _n$ (different plot symbols) in MEG data (green symbols; average over subjects) and in the model simulated at baseline parameters with $K = 100$ subsystems of $n_{\rm sub} = 100$ neurons each (violet symbols).  Distributions  for different $\epsilon$ collapse onto a single non-exponential master curve for both data and model. Corresponding distribution in phase-scrambled MEG signals   shows an exponential behavior, with absence of high excitation events (brown = surrogate data). \textcolor{black}{Inset: Rescaled $P(A_\epsilon)$ (green = average over subjects; violet = average over model simulations) and respective standard deviation (colored areas) shown for bin size  $\epsilon = 2$T.}  (D) Rescaled distributions of quiescence durations, $P(I_{\epsilon})$ collapse onto a single master curve for different $\epsilon$. Plotting conventions and model simulation details are the same as in (C).  \textcolor{black}{Top inset: Rescaled $P(I_\epsilon)$ (green = average of subjects; violet = average over model simulations) and respective standard deviation (colored area) shown for bin size  $\epsilon = 2$T}. Bottom inset: Probability $P_0$ of finding a quiescent time bin scales approximately as $P_0 = \exp\left(-a\epsilon^{\beta_I}\right)$ with bin size $\epsilon$; $\beta_I = 0.582 \pm 0.013$ and $\beta_I = 0.610 \pm 0.012$ for data and model, respectively; $\beta_I = 0.996 \pm 0.001$ for surrogate data.}
\label{fig:3}
\end{figure*}

We start by defining the instantaneous network excitation, $A_{\epsilon}(t)$,  as the number of extreme events co-occurring within  time bins of size $\epsilon$ across the entire MEG sensor array (Appendix D).
For each sensor, extreme events are the extreme points in that sensor's signal that exceed a set threshold $e$~(Fig.~\ref{fig:3}A). For a given threshold, network excitation $A_{\epsilon}$ depends on the size of the time bin $\epsilon$ that we use to analyze the data~(Fig.~\ref{fig:3}B).
\textcolor{black}{To make contact with the model, we parcel our simulated network  into $K$ equally-sized disjoint subsystems of $n_{\rm sub} = N/K$ neurons each, and consider each subsystem activity $m_\mu$, $\mu=1,\dots,K$, as the equivalent of a single MEG sensor signal. Network excitation, $A_{\epsilon}$, for the model then follows the same definition as for the data, allowing us to perform direct side-by-side comparisons of extreme event statistics.} 

 We first study the distribution of network excitation, $P(A_{\epsilon})$. We set $e = 2.9$SD both for MEG  data and for the model \cite{shriki13}. Even though $P(A_{\epsilon})$ generally depends on  $\epsilon$, the distributions corresponding to different $\epsilon$ collapse onto a single, non-exponential master curve when $A_{\epsilon}$ is rescaled by  $\langle A_{\epsilon}\rangle$,  the average instantaneous network excitation~(Fig.~\ref{fig:3}C).  Excitation distribution is thus invariant under temporal coarse-graining  and  the number of extreme events scales non-trivially  with $\epsilon$, in contrast to phase-shuffled surrogate data (Appendix E).
Remarkably,  model simulations fully recapitulate this data collapse as well as the non-exponential extreme event statistics. 

\textcolor{black}{Model simulations reproduce the distribution of network excitation, $P(A_\epsilon)$, to within the variability observed among subjects~(Fig.~\ref{fig:3}C, inset), for given values of $\epsilon$.   To quantify  how close the model distribution $P_{\rm m}(A_{\epsilon})$ is to the data-derived average distribution $P_{\rm d}(A_{\epsilon})$, we calculate the  Kullback–Leibler (KL) divergence~\cite{kl_divergence}:
\begin{equation}
\begin{aligned}
    D_{\rm dm}(A_{\epsilon}) \equiv & D_{KL}(P_{\rm d}(A_\epsilon)||P_{\rm m}(A_\epsilon)) = \\
    & \sum_{A_\epsilon}P_{\rm d}(A_\epsilon)\ln\frac{P_{\rm d}(A_\epsilon)}{P_{\rm m}(A_\epsilon)}.
\end{aligned}
\label{eqn:kld}
\end{equation}
This is to be compared with  the average KL divergence across subjects:  $D_{\rm dd}(A_{\epsilon}) = \langle D_{KL}(P_{\rm d}^i(A_\epsilon)||P_{\rm d}^j(A_\epsilon)) \rangle_{i,j}$, averaged across all pairs of MEG subjects indexed by $i$ and $j$. The data-model divergence is very small  and within the range of variability across subjects ($D_{\rm dm}(A_\epsilon) \lesssim D_{\rm dd}(A_\epsilon)$, see Table \ref{fig:5}), suggesting that the model quantitatively reproduces the measured distributions to the degree that can be expected given natural variability in the data.}

Periods of excitation ($A_\epsilon \neq 0$) are separated by  periods of quiescence ($A_\epsilon = 0$) of duration $I_{\epsilon} = n \epsilon$, where $n$ is the number of consecutive time bins with $A_{\epsilon} = 0$. The distribution of quiescence durations, $P(I_{\epsilon})$, is invariant under temporal coarse-graining when rescaled by $\langle I_{\epsilon} \rangle$, the average quiescence duration, collapsing onto a single, non-exponential master curve~(Fig.~\ref{fig:3}D). \textcolor{black}{As was the case with the distribution of network excitation, the model-predicted distribution of quiescence durations, $P_{\rm m}(I_\epsilon)$, also diverges from the data average $P_{\rm d}(I_\epsilon)$ by an amount that is within the range of variability among subjects~(Fig.~\ref{fig:3}D, inset; Table 1).}

We furthermore show that the overall probability $P_0(\epsilon)$ of finding a quiescent time bin follows a non-exponential relation $P_0(\epsilon)=\exp\left(- a \epsilon^{\beta_I}\right)$, with $\beta _I \simeq 0.6$~(Fig.~\ref{fig:3}D, lower inset), indicating that extreme events grouped into bins of increasing size are not independent \cite{meshulam2019}. 
These results are robust to changes in $N$ and $n_{\rm sub}$ so long as the number of subsystems $K$  is fixed or does not change considerably~(Figs S9--S10); otherwise, the threshold $e$ that defines an extreme event should be adjusted accordingly~(Fig.~S11), in particular, to closely reproduce the distribution of quiescence durations $P(I_{\epsilon})$~(Fig.~S11). \textcolor{black}{Finally we notice that the quantities $\langle A_{\epsilon} \rangle$ and $\langle I_{\epsilon} \rangle$ scale as a power of the bin size $\epsilon$ (Fig.~S12), and are connected to each other by a relationship of the form $\langle A_{\epsilon} \rangle \sim \langle I_{\epsilon} \rangle^{b_{AI}}$ (Fig.~S12). This implies that, for a fixed value of $e$, both the distribution  $P(A_{\epsilon})$ and $P(I_{\epsilon})$ are controlled by a single quantity, e.g., the average network excitation $\langle A_{\epsilon} \rangle$.}

In sum, our simple model at baseline parameters provides a robust account of the collective statistics of extreme events. We emphasize that the excellent match to the observed long-tailed distributions is only observed for the inferred value $\beta\simeq 0.99$ very close to criticality; already for $\beta=0.98$, we observe significant deviations from data ~(Figs.~S13, S14), demonstrating that excitation and quiescence distributions represent a powerful benchmark for collective brain activity.

\subsection{Scale-free neuronal avalanches occur concomitantly with oscillations}

\begin{figure}
\centering
\includegraphics[width=1\linewidth]{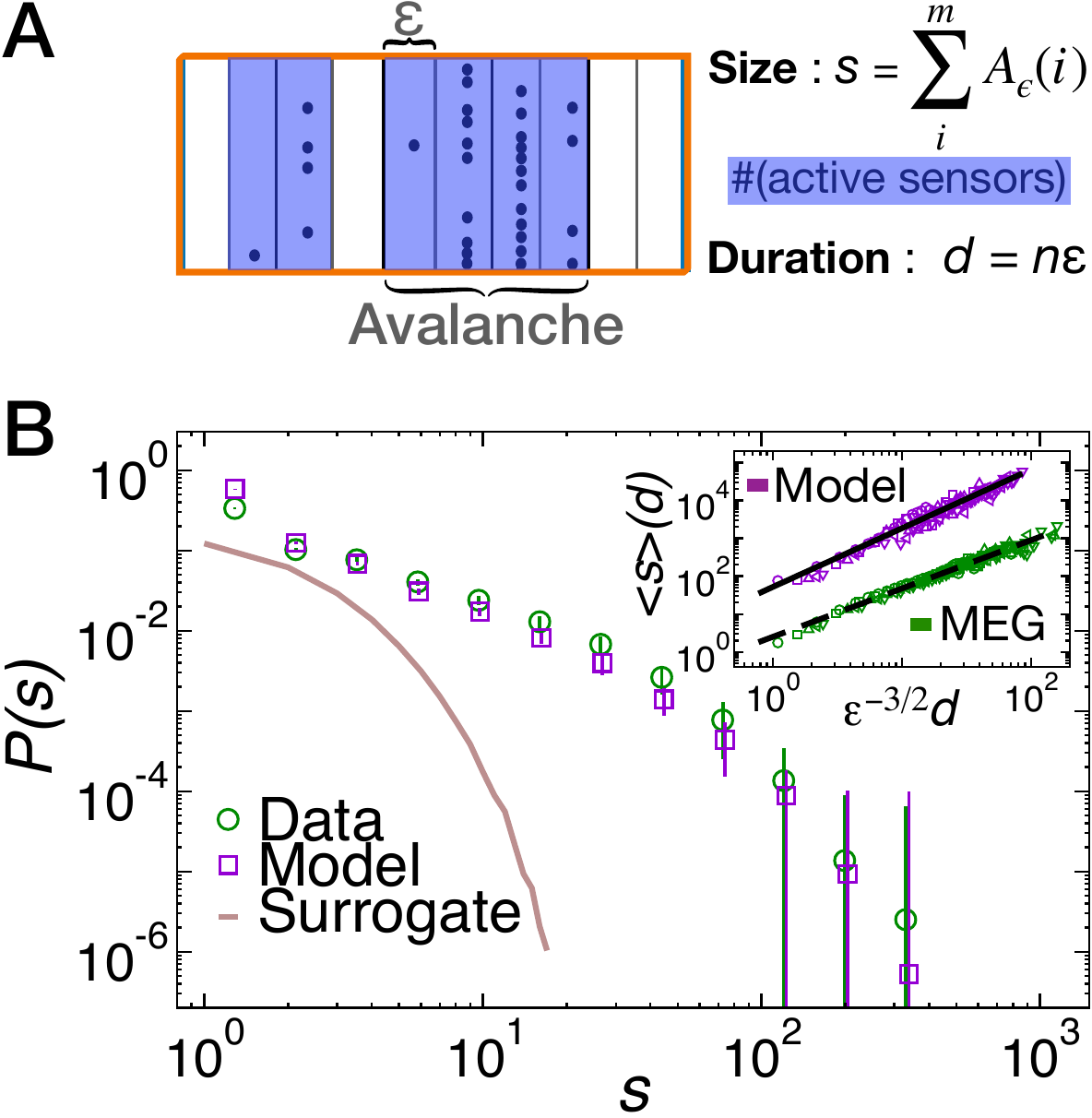}
\caption{Scale-free neuronal avalanches in MEG resting state activity are reproduced by a marginally subcritical adaptive Ising model. (A) Schematic representation of a neuronal avalanche. Avalanche size $s$ is the sum of network excitations $A_\epsilon$ over time bins belonging to the avalanche; its duration $d$ is the number of bins times their duration, $\epsilon$. (B) Distribution of avalanche sizes, $P(s)$, for MEG data (green circles with error bars = average over subjects $\pm$ standard deviation) and the model simulated at baseline parameters with $K = 100$ subsystems of $n_{\rm sub} = 100$ neurons each (violet squares with error bars = average over model simulations $\pm$ standard deviation). Both distributions are estimated using a threshold $e = 2.9$SD and bin size $\epsilon_4=4T$. The brown curve is the distribution $P(s)$ obtained from surrogate data (Appendix E) with the same threshold and bin size. 
Inset: Average avalanche size scales with its duration as $\langle s \rangle \sim d^\zeta$ (different plot symbols = different $\epsilon$ as in Fig.~\ref{fig:3}; green = MEG data; violet = model simulation; model simulation curves are vertically shifted for clarity), so that the exponent $\zeta$ remains independent of the time bin size $\epsilon$. $\zeta = 1.28\pm 0.01$ for MEG data (dashed line) and $\zeta = 1.58 \pm 0.03$ for model simulation (thick line).}
\label{fig:4}
\end{figure}

A neuronal avalanche is a maximal contiguous sequence of time bins populated with at least one extreme event per bin~(Fig.~\ref{fig:4}A)~\cite{plenz:pl03,shriki13}; every avalanche thus starts after and ends with a quiescent time bin  ($A_\epsilon=0$) (see Appendix D for details). 
Typically, neuronal avalanches are characterized by their size $s$, defined as the total number of extreme events within the avalanche. Avalanche sizes have been reported to have a scale-free power-law distribution~\cite{plenz:pl03,plenz:awa,shriki13,fontanele_2019_prl,fl2020_lrtc}. 

We estimate the distribution $P(s)$ of avalanche sizes in the resting state MEG, and compare it with the distribution obtained from model simulation at close-to-critical baseline parameter set~(Fig.~\ref{fig:4}B).  Both distributions are described by a power-law with an exponential cutoff~\cite{shriki13} and show an excellent match across subjects and for individual subjects. \textcolor{black}{Again, the KL divergence between the mean empirical and the model  distribution is smaller than the mean KL divergence estimated among MEG subjects~(Table 1).} Phase scrambled surrogate data strongly deviate from the power-law observations, as do model predictions when parameter $\beta$ is moved even marginally below $0.99$~(Fig.~S15). These results are independent of the $N$ and $n_{\rm sub}$ so long as the number of subsystems $K$  is fixed (Fig.~S16) or does not change considerably~(Fig.~S11).
%
Importantly, the model also reproduces the scaling relation $\langle s\rangle(d) \sim d^{\zeta}$ that connects average avalanche sizes $s$ and durations $d$. Unlike the power-law exponent of avalanche size distribution that typically depends on time bin size $\epsilon$~\cite{plenz:pl03,fl2020_lrtc}, the exponent $\zeta$  does not depend on $\epsilon$, as shown by the data collapse for both MEG data and model (Fig.~\ref{fig:4}B, the inset). While the scaling behavior is reproduced qualitatively, the  inferred and model-derived values of $\zeta$ are not in quantitative agreement, likely due to the overly simplified mean-field connectivity assumed by our model.  

\renewcommand{\figurename}{Table}
\setcounter{figure}{0}
\begin{figure}
\centering
\includegraphics[width=1\linewidth]{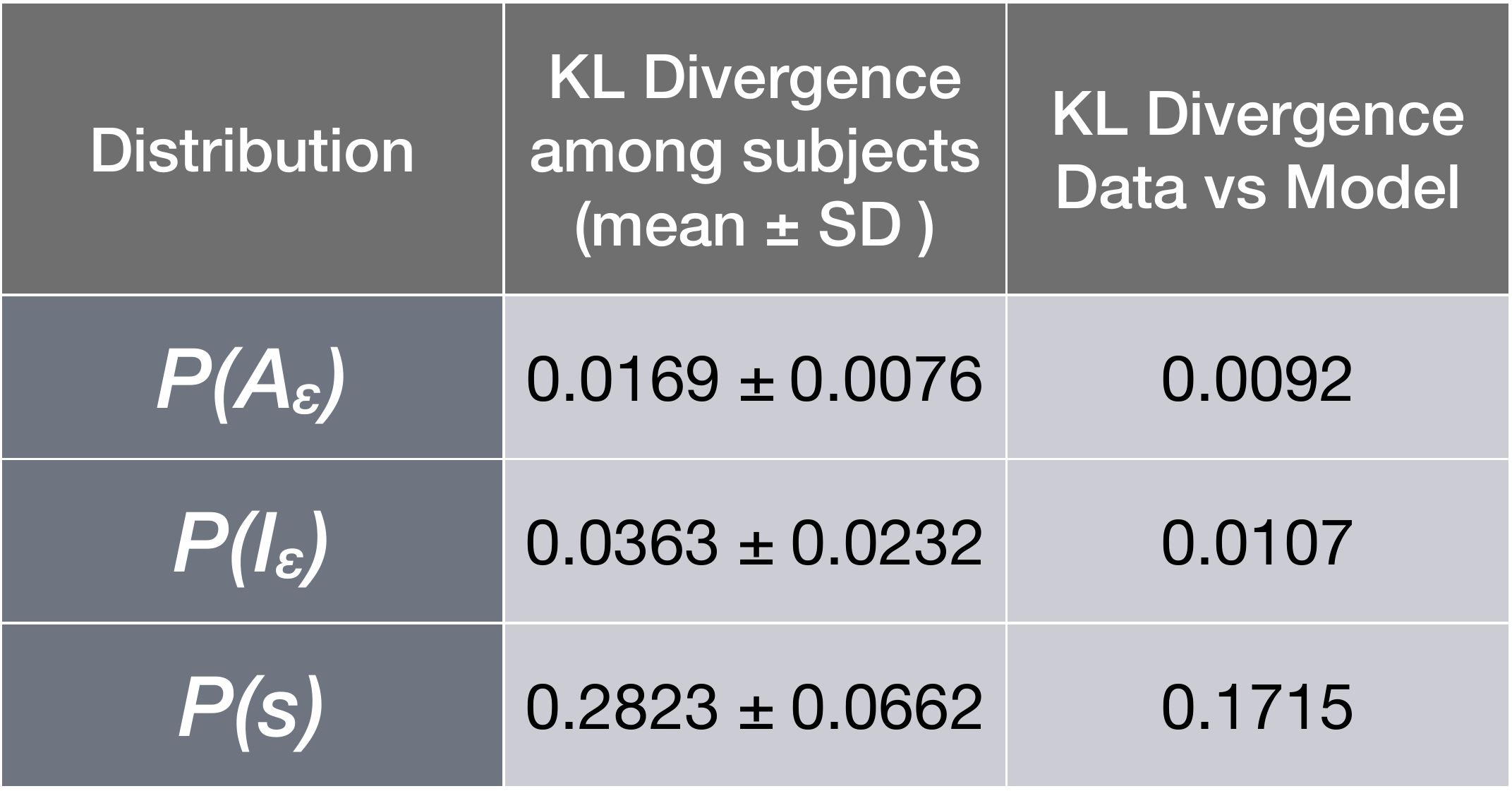}
\caption{The adaptive Ising model that reproduces signal autocorrelation on an individual MEG sensor makes quantitative predictions about the distribution of network excitation [$P(A_\epsilon)$], the distribution of quiescence durations [$P(I_\epsilon)$], and the distribution of avalanche sizes [$P(s)$], collective quantities defined over the entire MEG sensor array. Average ($\pm$ SD) Kullback-Leibler (KL) divergence across all pairs of subjects (second column) quantifies the biological variability in these distributions. The mismatch between model-predicted and data-derived (across-subject-average) distributions is also quantified by the KL divergence and reported in the third row~(computed as in Eq~(\ref{eqn:kld})). }
\label{fig:5}
\end{figure}

\section{Discussion}
In this paper, we put forward the adaptive Ising class of models for capturing large scale brain dynamics. Quite generally, these models combine a microscopic and stochastic description of excitatory neuron spiking with coarse-grained mean-field model of activity-dependent feedback. This endows the models with several unique characteristics that we discuss in turn: \emph{(i)} the ability to generate a diverse range of stylized behaviors observed in real brain dynamics and locate these behaviors in the model's phase diagram; \emph{(ii)} the possibility to connect the model to a wide range of known theoretical results in statistical physics and theoretical neuroscience; \emph{(iii)} the ability to rigorously infer model parameters from data, quantitatively test its predictions across a range of spatial and temporal scales, and thus derive biological insight about brain function.

\subsection{Diversity of dynamical behaviors in a simple non-equilibrium model}  By combining  local interactions with a  time-varying field in the form of an activity-dependent feedback,  the  adaptive Ising model exhibits a phase transition to a self-oscillatory behavior at the Andronov-Hopf bifurcation critical point that inherits the characteristics of the classical Ising ferromagnet second-order phase transition~\cite{ddemartino_jphys_2019}. In proximity and slightly below the critical point, the ongoing network activity $m$ --- the equivalent of the Ising magnetization --- shows intermittent oscillations, whose associated reversal time and zero crossing area are power-law distributed. To our knowledge, this is the simplest model class that reproduces the stylized coexistence of neuronal avalanches and oscillations, the two antithetic features of real brain dynamics. Moreover, these features coexist already in the most basic, mean-field formulation of the model, whose phase diagram can be computed analytically. Interestingly, in this formulation, individual units are neither intrinsic oscillators themselves~\cite{deco_2017_scirep,buendia_2021_prr}, nor are they mesoscopic units  operating close to a Hopf bifurcation~\cite{freyer_2012_jneuro,deco_2017_scirep,cabral_2017_neuroim}, and the collective dynamics is therefore not a result of oscillator synchronization (even though this regime can be captured as well by a different realisation of an adaptive Ising model). Our proposal thus provides an analytically-tractable alternative to---or perhaps a reformulation of---existing models~\cite{poil_2012_jneuro, disanto_2018_pnas}, which typically implicate particular excitation/inhibition or network resource balance to open up the regime where oscillations and avalanches may coexist. 

\subsection{Connections to results in statistical physics}  Starting with the seminal work of Hopfield~\cite{hopfield1982neural}, the functional aspects of neural networks have traditionally been studied with microscopic spin models or attractor neural networks. These systems qualitatively reproduce  the associative memory functionality ascribed to real neural networks~\cite{amit1992modeling} and their phase transitions have been thoroughly studied with statistical mechanics tools ~\cite{amit1985spin}. The associated inverse (maximum entropy) problem recently attracted great attention in connecting spin models to data~\cite{schneidman2006weak,tkavcik2014searching}, in particular with regard to criticality  signatures~\cite{tkavcik2015thermodynamics}. \textcolor{black}{While the initial formulations of the maximum-entropy problem aimed at modeling the stationary distribution from which activity patterns were drawn independently, later work specifically focused on the temporal correlation structure in the neural  activity, by modeling across-time interactions between individual neurons~\cite{tyrcha2013,marre2009prediction,nasser2013spatio}. Parallel work meanwhile considered spin-models responding to global known modulation, e.g., via stimulus, in the stimulus-dependent maximum-entropy approach~\cite{granot2013stimulus}, or global latent modulation which can result in critical (Zipf-like) stationary distributions of activity~\cite{schwab2014zipf,aitchison2016zipf,humplik2017probabilistic}, or dynamics reminiscent of avalanches~\cite{mora2015dynamical}. }

\textcolor{black}{Despite such generalizations, the dynamical expressive power of maximum-entropy stationary, kinetic, or latent-variable models has been limited: while detailed patterns of short-term temporal correlations and statistical criticality could be captured from data, the rhythmic behavior of brain oscillations was beyond the practical scope of these models. Adaptive Ising model class can be seen as a natural, yet orthogonal, extension to the previous work that enables oscillations and furthermore permits us to explore an interesting interplay of mechanisms, for example, by having self-feedback drive Hopfield-like networks (with memories encoded in the coupling matrix $J$) through sequences of stable states. }

Another link to rich existing literature becomes apparent when we consider connectivity degrees of freedom (encoded in $J$ and in the pooling that drives the feedback; Appendix B) in order to model cell types (e.g., inhibitory vs excitatory), the spatial structure of the cortex, or to capture empirically established topological features of real neural networks. Within equilibrium statistical mechanics, the effects of various lattices or disordered connectivity in spin systems have been studied rigorously; phenomenological arguments, supported by preliminary results,  suggest that adding the feedback that drives the adaptive Ising model out of equilibrium does not change the fundamental nature (e.g., critical exponents) of the critical point. This permits us to directly translate existing equilibrium results into the non-equilibrium setup: an example would be the introduction of scale-free topology among excitatory neurons into our model (example in the Appendix B). Looking towards the future, powerful tools of statistical physics, such as the Landau-Ginzburg theory and the renormalization group, can be brought to bear on the generalizations of the non-equilibrium adaptive Ising model, enabling rapid progress based on  existing equilibrium results.

\subsection{Inferring the model to derive biological insights}
Since the adaptive Ising model can be solved analytically in the mean field limit, we can infer its two key parameters by matching the auto-correlation function of MEG signals to the model-predicted auto-correlation function. In contrast to previous work~\cite{disanto_2018_pnas,poil_2012_jneuro}, we do not make contact with existing data by qualitatively matching the phenomenology, but by proper parameter inference. The inferred parameters consistently place the  model very close to its critical point, supporting the hypothesis that alpha oscillations represent brain tuning to criticality \cite{linken01,freyer_2009,linken12,freyer_2012_jneuro,palva13}.  The possibility of mapping  empirical data to a defined region in the adaptive Ising model phase diagram through parameter inference paves the way for further quantification of the relationship between measures of brain criticality and healthy, developing, or pathological brain dynamics along the lines developed  recently~\cite{fakete2021scirep,palva2015jn,oshrit2016jn}.
 
Our inferred model makes a wide range of further predictions that can be confronted with data. By parcelling the simulated system into groups of spins, we can mimic the signals captured by multiple MEG sensors over the cortex. We find a remarkable quantitative correspondence between the non-exponential and scale-invariant distribution of network excitation observed across the MEG sensors and in our simulation, which holds when averaged across individuals or within single individuals. 
Our analysis clearly demonstrates that extreme events are non-independent across space and time. This non-trivial spatio-temporal organization of extreme events is strongly indicative of a network state close to criticality~\cite{taglia_2012_point,fraiman12}; the agreement between model and data breaks down for surrogate data as well as for models further removed from criticality.  Moreover, the extreme events coalesce into neuronal avalanches as reported previously~\cite{plenz:pl03,plenz:awa,shriki13,fl2020_lrtc}, which our model reproduces as well. Taken together, our model provides a remarkably broad account of brain dynamics across spatial and temporal scales. 
 
 \textcolor{black}{Despite these successes, we openly acknowledge the quantitative failures of our model: \emph{(i)} at the single sensor level, small deviations exist in the distributions of log activity~(Fig.~\ref{fig:2}C), likely due to very long timescales or  non-stationarities in the MEG signals~\cite{shriki13,stam1999alpha_nonlin,stam2005nonlin}; \textcolor{black}{\emph{(ii)} small deviations  beyond the range of data variability exist in the probability distribution $P(I_{\epsilon})$ for a narrow range of intermediate quiescence intervals, even when the rest of the distribution is reproduced very well~(Fig.~\ref{fig:3}D, inset);} \emph{(iii)} the scaling exponent governing the relation between the avalanche size and duration, $\zeta$, is not reproduced quantitatively~(Fig.~\ref{fig:4}B, inset).} Furthermore, beyond the two key model parameters that were inferred directly from individual sensors ($\beta,c$), quantitative data analysis of extreme events requires additional parametric choices (time bin $\epsilon$, threshold $e$, system size $N$ and subsystem size $n_{\rm sub}$), both for empirical data as well as model simulations. While we successfully demonstrate the scaling invariance of the relevant distributions with respect to $\epsilon$ and robustness with respect to $N$ and $n_{\rm sub}$ at fixed $N/n_{\rm sub}$, a close match to data still requires choosing one extra parameter (e.g., threshold $e$). \textcolor{black}{Concerning this point, we have verified that an even closer agreement between empirical and numerical distributions can be achieved setting  slightly different threshold values $e$ on MEG and model data---in particular, when the thresholds are adjusted so that the data and the model are perfectly matched in the average activity rate.}
 
 Despite these valid points of concern, we find it remarkable that such a simple and tractable model can quantitatively account for so much of the observed phenomenology. \textcolor{black}{Future work should first consider connectivity beyond the simple all-to-all mean-field version that we introduced here, likely leading to a better data fit and new types of dynamics, e.g, cortical waves.}  \textcolor{black}{Preliminary model simulations show that the exponent $\zeta$, which connects avalanche sizes and durations, is affected by the connectivity and, furthermore, more closely matches the value of $\sim 1.3$ characteristic of the data for nearest-neighbor 2D connectivity on a square lattice.} 
 \textcolor{black}{Second, we strongly advocate for rigorous and transparent data analysis and quantitative---not only stylized---comparisons to data. To this end, care must be taken not only when inferring the essential model parameters  beyond the linear approximation~\cite{giardina2020prx}, but also when treating the hidden ``degrees of freedom'' related to data analysis (specifically,  subsampling, temporal discretization, thresholding etc.)~\cite{priesemann2009,levina2017sub,plenz:pl03,fl2020_lrtc,shriki13}.} Third, it is important to confront the model with different types of brain recordings; a real success in this vein would be to account simultaneously for the activity statistics at the microscale (spiking of individual neurons) as well as at the mesoscale (coarse-grained activity probed with MEG, EEG, or LFP). 

\section*{Acknowledgements}
FL acknowledges support  from the European Union's Horizon 2020 research and innovation program under the Marie Sklodowska-Curie Grant Agreement No. 754411. GT acknowledges the support of the Austrian Science Fund (FWF) under Stand-Alone Grant No. P34015.

\section*{Appendix A: Data acquisition and pre-processing}
Ongoing brain activity was recorded from 14 healthy  participants in the MEG core facility at the NIMH (Bethesda, MD, USA) for a duration of 4 min (eyes closed). All experiments were carried out in accordance with NIH guidelines for human subjects. The sampling rate was 600 Hz, and the data were band-pass filtered between 1 and 150 Hz. Power-line interferences were removed using a 60 Hz notch filter designed in Matlab (Mathworks). The sensor array consisted of 275 axial first-order gradiometers. Two dysfunctional sensors were removed, leaving 273 sensors in the analysis. Analysis was performed directly on the axial gradiometer waveforms. The data analyzed here were selected from a set of MEG recordings for a previously published study \cite{shriki13}, where further details can be found. For the present analyses we used the subjects showing the highest percentage of spectral power in the alpha band (8-13 Hz). Similar results were obtained for randomly selected subjects.

\section*{Appendix B: Further details on the Adaptive Ising model}
The model is composed of a collection of $N$ spins $s_i = \pm 1$ ($i = 1,2,...,N$) that  interact with each other with a coupling strength $J_{ij}$. In our analysis, the  $N$ spins  represent excitatory neurons that are active when $s_i = + 1$ or inactive when $s_i = -1$, and $J_{ij} > 0$. Furthermore, we consider the fully  homogeneous scenario, with neurons interacting  with each other through  synapses of equal strength $J_{ij} = J = 1$. However,  interesting generalization  with non-homogeneous, negative, non-symmetric $J_{ij}$ are possible, to include, for example, the effect of inhibitory neuronal population and   structural and functional heterogeneity.  
The $s_i$ are stochastically activated according to the Glauber dynamics, where the state of a neuron is drawn from the marginal Boltzmann-Gibbs distribution 
\begin{equation}
P(s_i) \propto \exp(\beta \tilde{h}_is_i) \quad \tilde{h}_i = \sum_j J_{ij} s_j + h_i
\end{equation}
The spins experience an external field $h$, a negative feedback that depends on network activity according  to the following equation, 
\begin{equation}
\dot{h}_i = -c \frac{1}{|\mathcal{N}_i|}\sum_{j\in \mathcal{N}_i}  ^{|N_i|} s_j,
\label{eq:feed}
\end{equation}
where $c$ is a constant that controls the feedback strength, and the sum runs over a neighborhood of the neuron $i$ specified by $\mathcal{N}_i$; index $j$ enumerates over all the elements of this neighborhood. Depending on the choice of $\mathcal{N}_i$, the feedback may depend on the activity of the neuron $i$ itself (self-feedback), its nearest neighbors, or the entire network --- the case we considered in the main paper. In a more realistic setting including both excitatory ($J_{ij} > 0$) and inhibitory neurons ($J_{ij} < 0$), one could then take into account the different structural and functional properties of excitatory and inhibitory neurons by considering different interaction and feedback properties \cite{bonifazi2009gabae}.
In our simulations, one  time step  corresponds to one system sweep  --- i.e. $N$ spin flips --- of Monte Carlo updates, and Eq~(\ref{eq:feed}) is integrated using  $\Delta t = 1/N$. Note that this choice of timescales for deterministic vs stochastic dynamic is important, as it interpolates between the quasi-equilibrium regime where spins fully equilibrate with respect to the field $h$, and the regime where the field is updated by feedback after each spin-flip and so spins can constantly remain out of equilibrium.    $\Delta t$ is generally much smaller than the characteristic time of the adaptive feedback that is controlled by the parameter $c$.

\subsection*{Mapping between the adaptive Ising model and an E-I network}

We first encode a classic E-I model that leads to sustained oscillations, into an Ising model framework. We can think of this model as a single network of $N$ units whose coupling matrix $J_{ij}$ is asymmetric and is structured into two blocks that correspond to an excitatory and an inhibitory subpopulation. Specificcally, the network consists of a population 1, which is self-exciting with strength $J_{11}$ and which  excites population 2 with strength $J_{12}$, while population 2 is inhibiting population 1 with strength $J_{21}$. 

%
It can be demonstrated that the mean-field  dynamics of this stochastic system of Ising-like neurons in the limit of large populations is described by a Liouville deterministic equation of the form ($m_i$ $i=1,2$ is the average spiking rate of population $i$)~\cite{buhmann1987noise}:
\begin{eqnarray}
\dot{m}_1 &=& -m_1 + \tanh(J_{11} m_1 + J_{21} m_2) \\
\dot{m}_2 &=& -m_2 + \tanh(J_{12} m_2) \\
\end{eqnarray}
  
The E-I network has an ergodic state where $(m_1=m_2=0)$. Stability analysis to small perturbations of this state reveals an Andronov-Hopf bifurcation towards self-oscillations, when $J_{11}=2$ and $-J_{12}J_{21} > 1$. Upon matching the coefficients of such a linear expansion:
\begin{eqnarray}
\ddot{m} + (1-\beta) \dot{m} + c \beta m =0 \\
\ddot{m}_1 + (2-J_{11}) \dot{m}_1 + (1-J_{11}-J_{12}J_{21}) m_1 =0 
\end{eqnarray} 
we get an approximate mapping into the parameters $\beta,c$ of the simplest adaptive Ising model:
\begin{eqnarray}
J_{11} =\beta +1 \\
J_{12}  = \sqrt {\beta (1+c)} \\
J_{21}  = -\sqrt {\beta (1+c)}  . 
\end{eqnarray}

\subsection*{The role of topology}
One of the most interesting questions about synchronization in neural networks is how general features of the interaction topology affect the collective behavior, i.e., how  structure affects  function in general terms~\cite{buzsaki2006rhythms}. 
From a modeling perspective, most  efforts have focused on studying the Kuramoto model (KM)~\cite{arenas2008synchronization}, where the individual  excitable units are already postulated to be  oscillators (for a discussion about this point see \cite{andronov2013theory}). Nevertheless, 
 no exact analytical results for the KM on general  networks are available up to now, with an intense debate currently focusing on the nature of the onset of synchronization in strongly heterogeneous topologies~\cite{peron2019onset}. 

In contrast, we provide here an heuristic argument that the adaptive Ising model directly inherits the wealth of knowledge accumulated about its equilibrium  counterpart, in particular, with regard to the features and the location of its critical point(s). The critical point characteristics have been {\em rigorously} determined in several geometries, from dimensional lattices to complex networks and small world~\cite{onsager1944crystal, baxter2016exactly, dorogovtsev2008critical}. 

The fact that equilibrium Ising results can be generalized to the adaptive case can be seen  directly from the application of the linear response theory, upon considering the Landau expression for the free energy: by construction, the bifurcation point of the dynamical model coincides with the critical point of the underlying equilibrium model.

For instance, for the case of uncorrelated tree-like random graphs, described by a degree distribution  $P(k)$~\cite{newman2018networks}, the linear response applied to a Curie-Weiss approximate expression for the free energy \cite{leone2002ferromagnetic}
leads to the following approximate dynamical equations for the firing rates of nodes with degree $k$, $m_k$ (where $\langle k \rangle$ is the mean degree, $\langle m \rangle =\sum_k P(k) m_k$ is the average firing rate and $\langle m_v \rangle = \sum_k \frac{k}{\langle k\rangle} P(k) m_k$ is the average firing rate upon following a random link):
\begin{eqnarray}
\dot{m}_k = -m_k +\tanh(\beta(k \langle m_v \rangle +h)) \quad  \forall k \\
\dot{h} = -c\langle m \rangle.
\end{eqnarray}

As it can be easily verified by linearizing around the stationary solution $m_k=h=0$, these equations show that the model has a bifurcation point located at the same position as  the equilibrium critical point, i.e. $(\beta J)_c=\frac{\langle k^2\rangle }{\langle k\rangle}$ (a more refined calculation \cite{leone2002ferromagnetic} based on asymptotically exact Bethe-Peierls approximation gives  $(\beta J)_c=-\frac{1}{2} \log (1-2\frac{\langle k\rangle }{\langle k^2\rangle})$).

This simple example shows that the inverse temperature gets renormalized by the branching ratio~\cite{newman2018networks} $\frac{\langle k^2\rangle }{\langle k\rangle}$, a topological measure of the density of links, or synaptic connections in our context,  that could be considered itself as the key control parameter driving the system in and/or out the synchronized phase.
A direct consequence for our case is that if the topology we were considering were scale free, i.e. with an heavy tail for the degree distribution  $P(k)\sim k^{-\gamma}$ $\gamma<3$, then $\beta_c \to 0$ and the system would always be in the synchronized phase, a feature shared by many collective phenomena in strongly heterogeneous networks~\cite{dorogovtsev2008critical}. In our case, subcritical dynamics is inferred from data and the scale-free topology is not appropriate, but the reasoning here demonstrates clearly how known facts about equilibrium Ising on different topologies can directly translate into the insights of the adaptive Ising model.

\section*{Appendix C: Detrended Fluctuations Analysis of the alpha band amplitude envelope} The DFA \cite{peng:1994} consists of the following steps: (i) Given a time series  $x_i (i = 1,...,N)$  calculate the integrated signal $I(k) = \sum_{i=1}^{k} (x(i) - \langle x\rangle)$, where $\langle x\rangle$ is the mean of $x_i$; (ii) Divide the integrated signal $I(k)$ into boxes of equal length $n$ and, in each box, fit $I(k)$ with a first order  polynomial $I_n(k)$, which represents the trend in that box; (iii) For each $n$, detrend  $I(k)$ by subtracting the local trend, $I_{n}(k)$, in each box and calculate the root-mean-square (r.m.s.)  fluctuation $F(n)= \sqrt{\sum_{k=1}^{N} [I(k)-I_{n}(k)]^2/N}$; (iv) Repeat this calculation over a range of box lengths $n$ and obtain a functional relation between $F(n)$ and $n$. For a power-law correlated time series, the  average r.m.s. fluctuation function $F(n)$ and the box size $n$ are connected  by a power-law relation, that is $F(n) \sim n^\alpha$. The exponent $\alpha$ quantifies the long-range correlation properties of the signal. Values of $\alpha < 0.5$ indicate the presence of anti-correlations in the time series $x_i$, $\alpha = 0.5$ absence of correlations (white noise), and $\alpha > 0.5$ indicates the presence of positive correlations in $x_i$. The DFA was applied to the alpha band ($8-13$ Hz) amplitude envelope. Data were band filtered in the range 8-13 Hz using a FIR filter (second order) designed in Matlab. The scaling exponent $\alpha$ was estimated in the $n$ range corresponding to 2s - 60s to avoid spurious correlations induced by the signal filtering \cite{linken01,linken12}.

\section*{Appendix D: Extreme events, instantaneous network excitation  $A_{\epsilon}$, neuronal  avalanches}


For each sensor, positive and negative excursions beyond a threshold $e$ were identified. In each excursion beyond the threshold, a single event was identified  at the most extreme value (maximum for positive excursions and minimum for negative excursions). Comparison of the signal distribution to the best fit Gaussian  indicates that the two distributions start to deviate from one another around $\pm 2.7$SD \cite{shriki13}. Thus, thresholds smaller than $\pm 2.7$SD will lead to the detection of many events related to noise in addition to real events whereas much larger thresholds will miss many of the real events. To avoid noise-related events while preserving most of relevant events, in this study with set the threshold $e$ at $\pm 2.9$ standard deviations (SD). The raster of identified events was binned at a number of temporal resolutions $\epsilon$, which are multiple of the sampling time $T = 1.67$ ms. The network excitation $A_{\epsilon}$ at a given temporal resolution $\epsilon$ is defined as the number of events occurring across all sensors in a time bin. An avalanche is defined as a continuous sequence of time bins in which there is at least an event on any sensor, ending with at least a time bin with no events (Fig.~\ref{fig:4}A). The size of an avalanche, $s$, is defined as the number of events in the avalanche. For further details see \cite{shriki13,fl2020_lrtc}.

\section*{Appendix E: Surrogate data and statistical analysis} Surrogate signals are obtained by random phase shuffling of the original continuous MEG signals. A Fourier transform of each sensor signal is performed, the corresponding phases are randomized  while amplitudes are preserved. The surrogate signals are then obtained by performing an inverse Fourier transform. The random phase shuffling destroys phase synchronization across cortical sites while preserving the linear properties of the original signals, such as power spectral density and two-point correlations \cite{theiler92}.\\
The $p-$value for the least square fit  performed in Figure 2F is the $p-$value of the coefficients. Error bars in the figures denote standard errors of the mean. 


\bibliography{bib_file_all.bib}

\clearpage

\onecolumngrid
\appendix

\section*{\Huge Supplementary Materials}

\vspace{2cm}

\renewcommand{\figurename}{Fig. S\hspace{-0.13cm}} 

\setcounter{figure}{0}

\begin{figure}[ht]
\centering
\includegraphics[width=\linewidth]{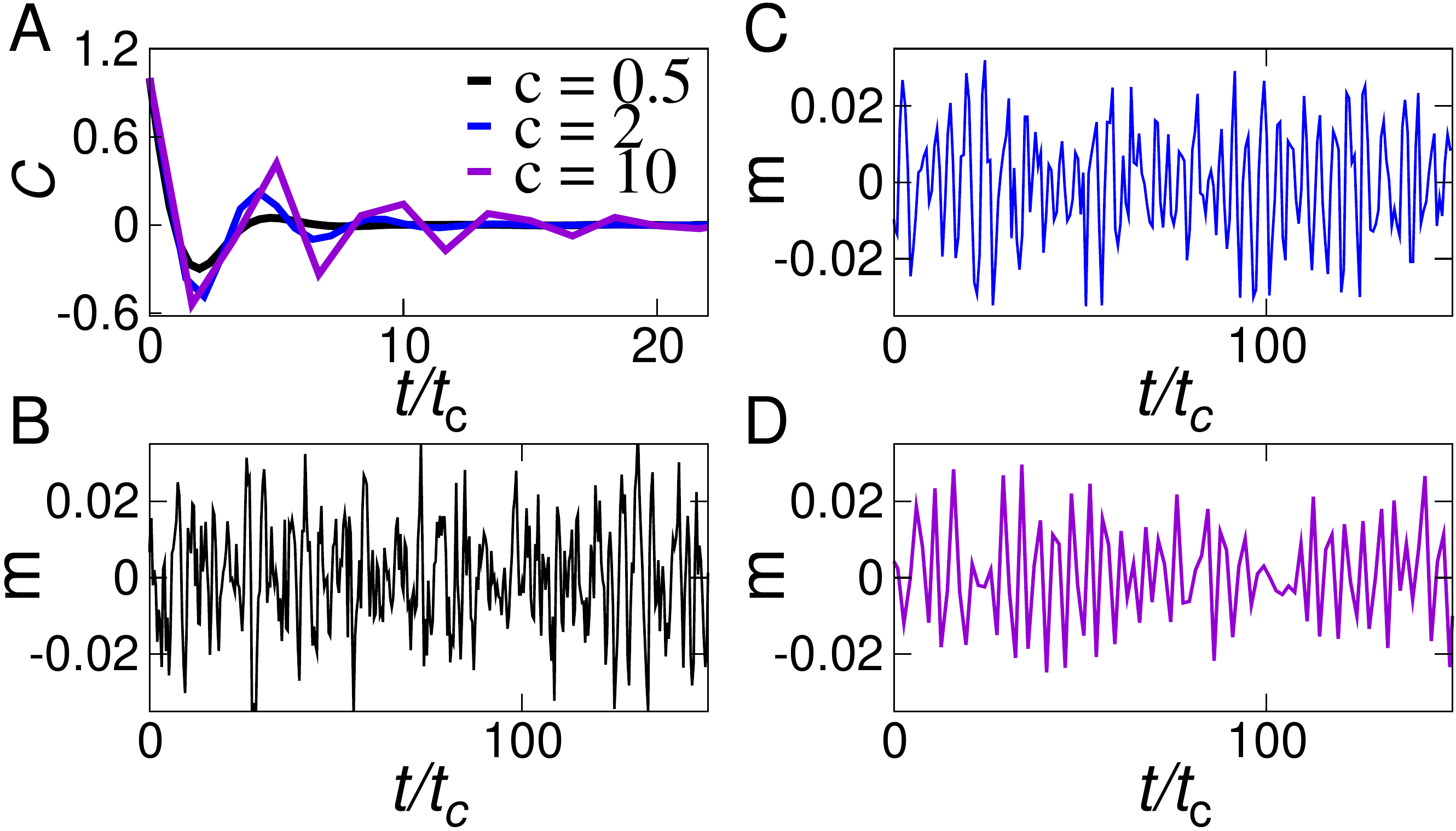}
\caption{{\bf Autocorrelation $C$ and ongoing network activity $m$ for $\beta = 0.5$ and different $c$ values.} Far from the critical point, the presence of a strong adaptive feedback may also produce short --- $C$ rapidly decays to zero --- intermittent oscillation bursts. (A) Autocorrelation for different $c$ values.  (B) $m$ for $c = 0.5$. (C) $m$ for $c = 2$. (D) $m$ for $c = 10$. $t_c$ is the inferred autocorrelation time  from exponential fit.}
\label{SI_fig:1}
\end{figure}

\begin{figure}[ht]
\centering
\includegraphics[width=\linewidth]{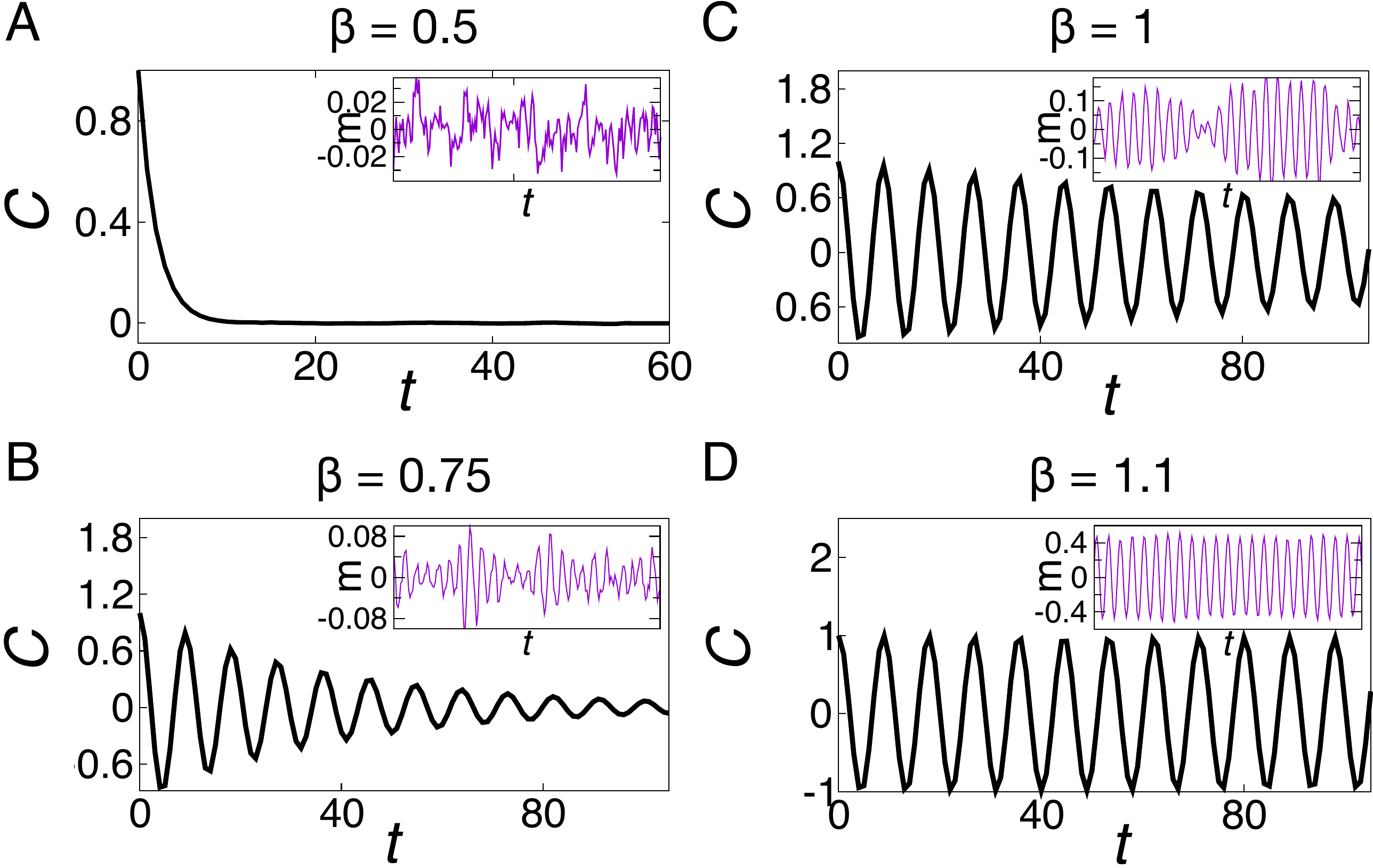}
\caption{{\bf The autocorrelation $C$ and ongoing network activity $m$ for $c = 0.5$ and  different values of the parameter $\beta$ controlling proximity to the critical/bifurcation point $\beta _c$.} In all cases the system is the resonant regime, and well above the transition line $c = c^*$. However, we observe that the system only develops consistent and structured  oscillations for large enough $\beta$ values, namely closer to the critical point $\beta = 1$. For $\beta > 1$, the system exhibits self-oscillations.}
\label{SI_fig:2}
\end{figure}

\clearpage

\begin{figure}[ht]
\centering
\includegraphics[width=\linewidth]{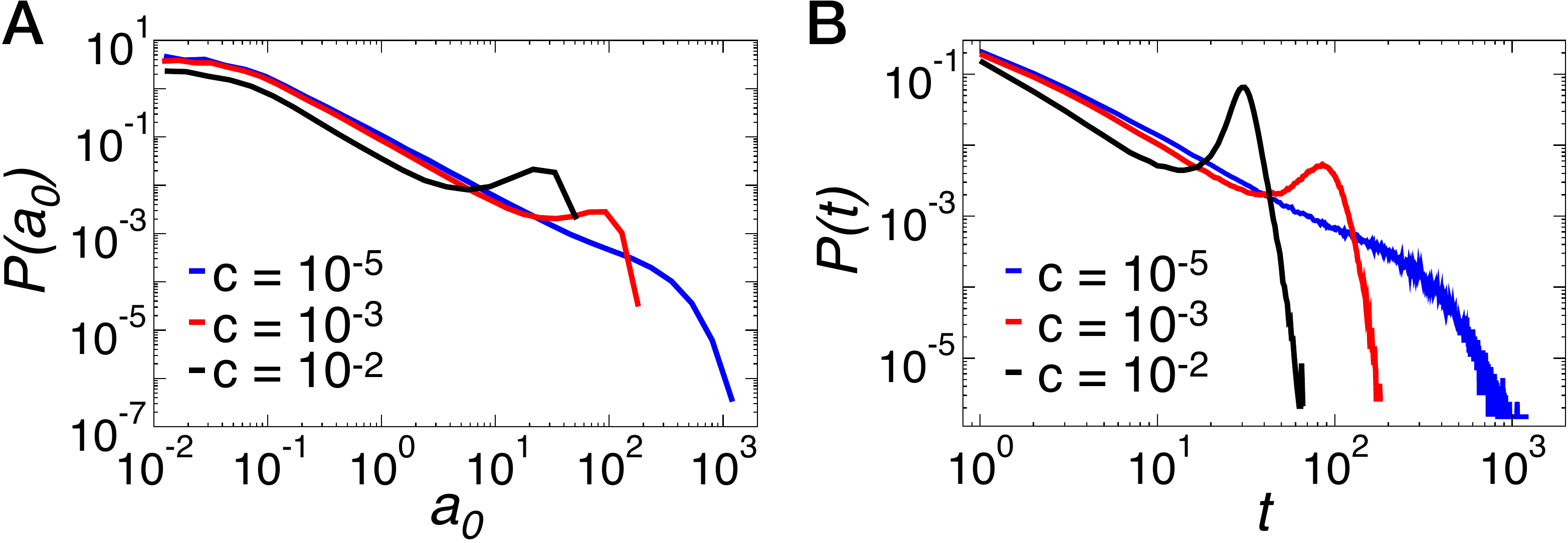}
\caption{{\bf The reversal time $t$ is defined as the time interval between consecutive  zero-crossing events in the ongoing network activity $m$ (Fig. 1). The quantity $a_0$ is the area under the curve between two zero-crossing events.} (A) Distribution $P(a_0)$ of the quantity $a_0$ for the model at the critical point $\beta = 1$ for the different strengths $c$ of the adaptive feedback. (B) Distribution $P(t)$ of the reversal time  for $\beta = 1$ and different  values of the parameter $c$.}
\label{SI_fig:3}
\end{figure}

\vspace{2cm}

\begin{figure}[hb]
\centering
\includegraphics[width=\linewidth]{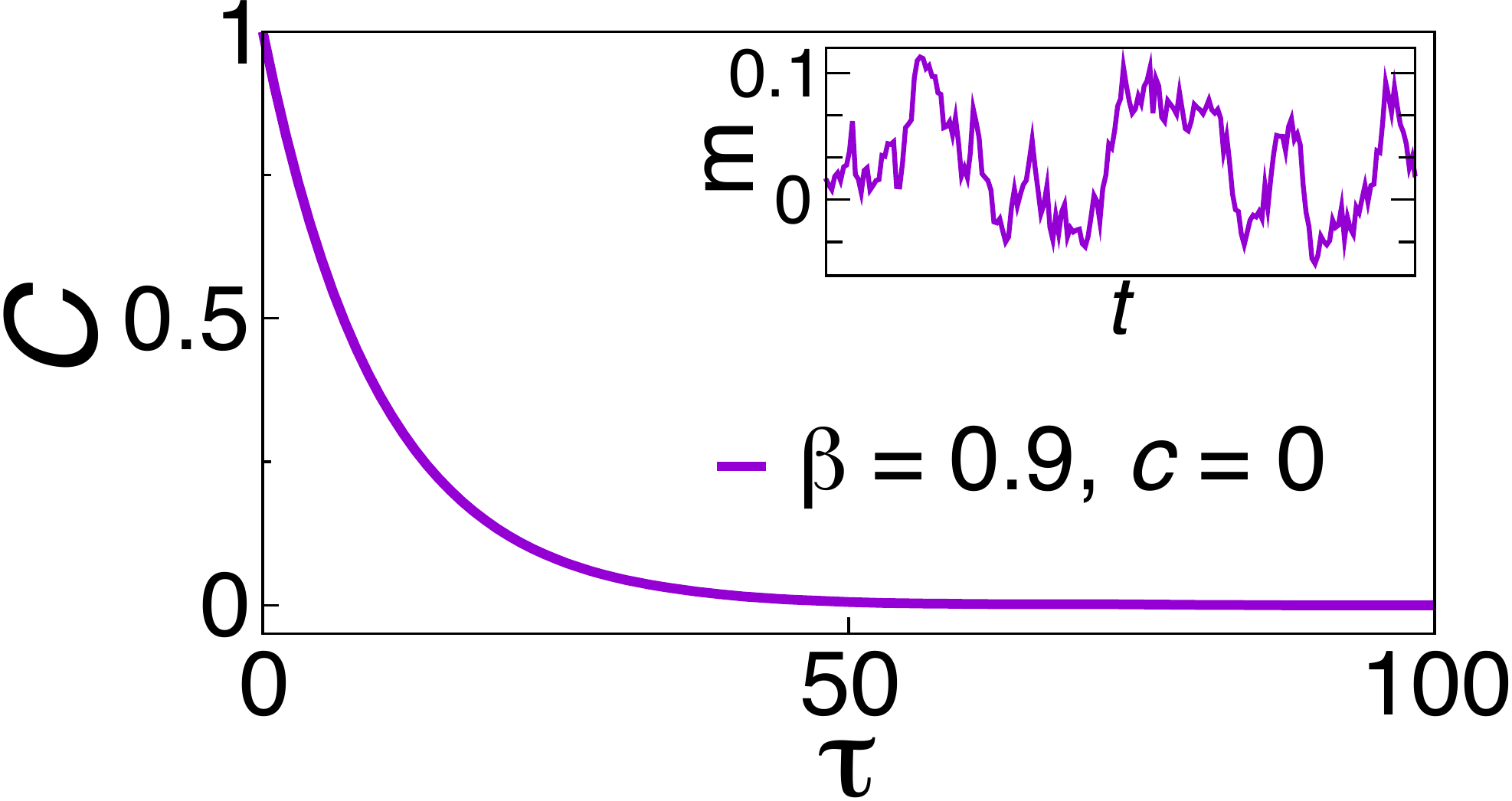}
\caption{{\bf The autocorrelation $C$ and ongoing network activity $m$ (inset) without adaptive feedback, i.e. $c = 0$.} Even though  $\beta$ is close to the critical value $\beta = 1$, the network does not exhibit any oscillatory behavior.}
\label{SI_fig:4}
\end{figure}

\begin{figure}[ht]
\centering
\hspace{-0.9cm}\includegraphics[width=0.95\linewidth]{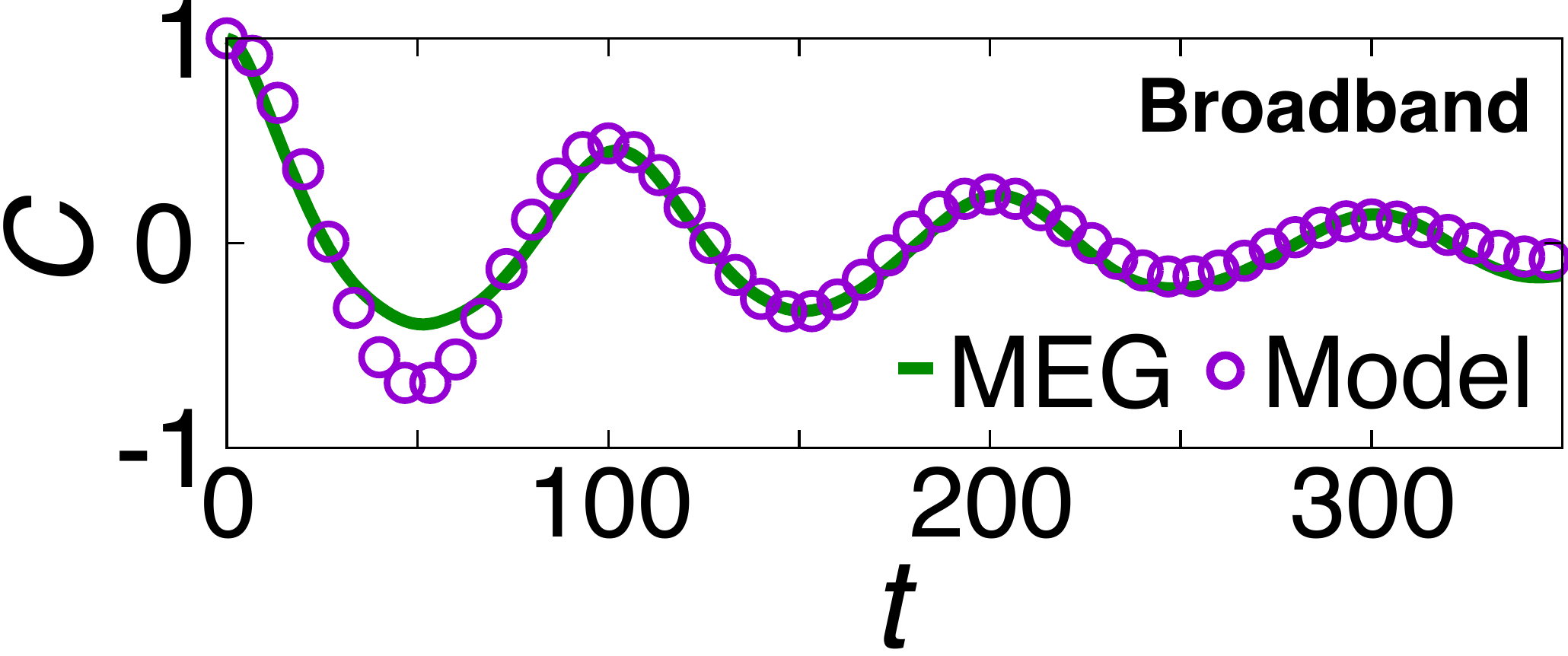}\vspace{0.5cm}
\includegraphics[width=\linewidth]{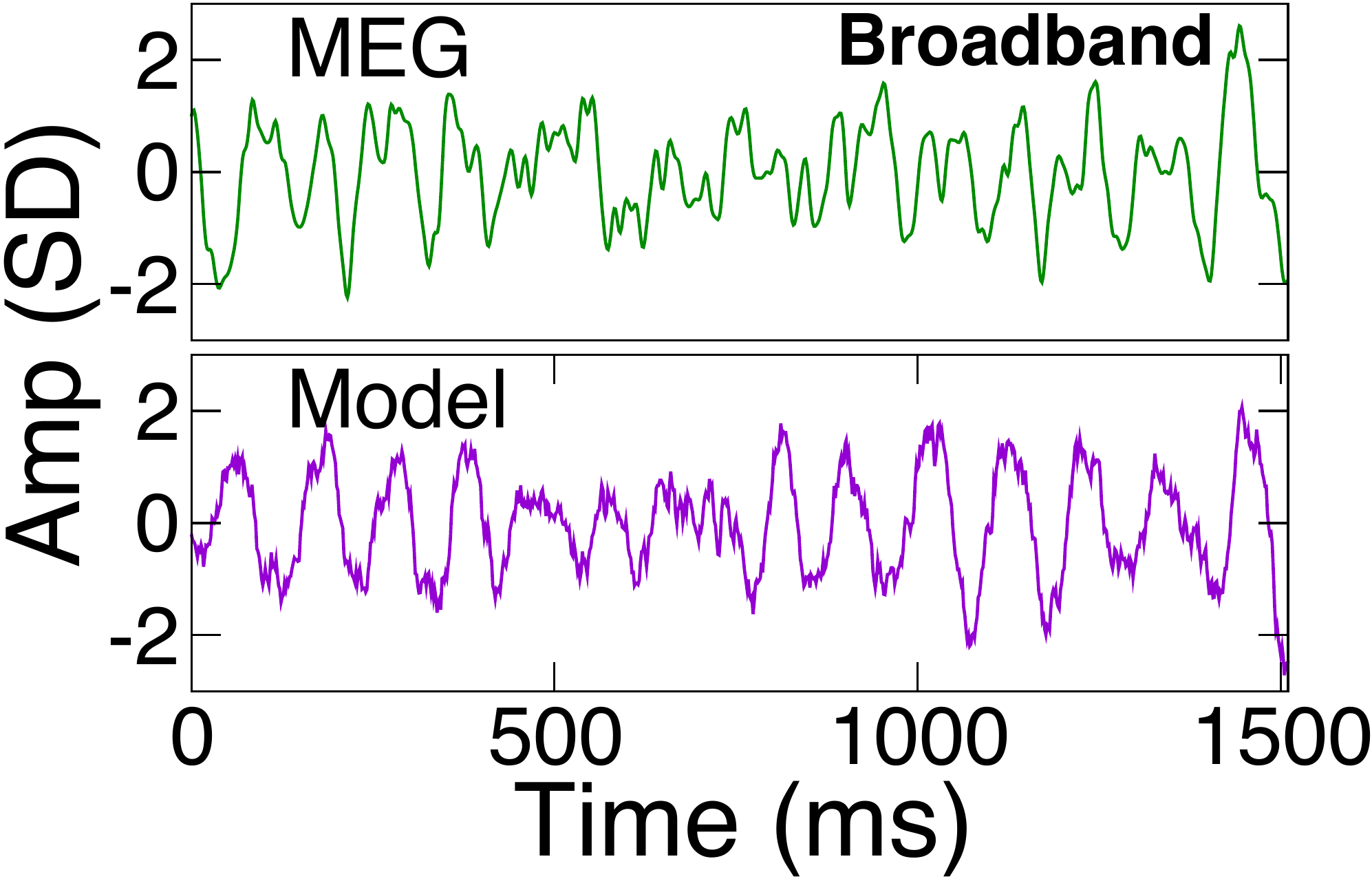}
\caption{{\bf Inference of network state from broadband signals.} (Top panel) Broadband MEG signal autocorrelation  and corresponding model autocorrelation for inferred parameters.  (Bottom panel) The broadband signal from a single MEG sensor in the resting awake state is compared with the network activity $m$ from model simulations with  parameters $\beta = 0.99$ and $c = 0.01$ inferred from the single sensor signal.}
\label{SI_fig:5}
\end{figure}

\begin{figure}[ht!]
\centering
\includegraphics[width=\linewidth]{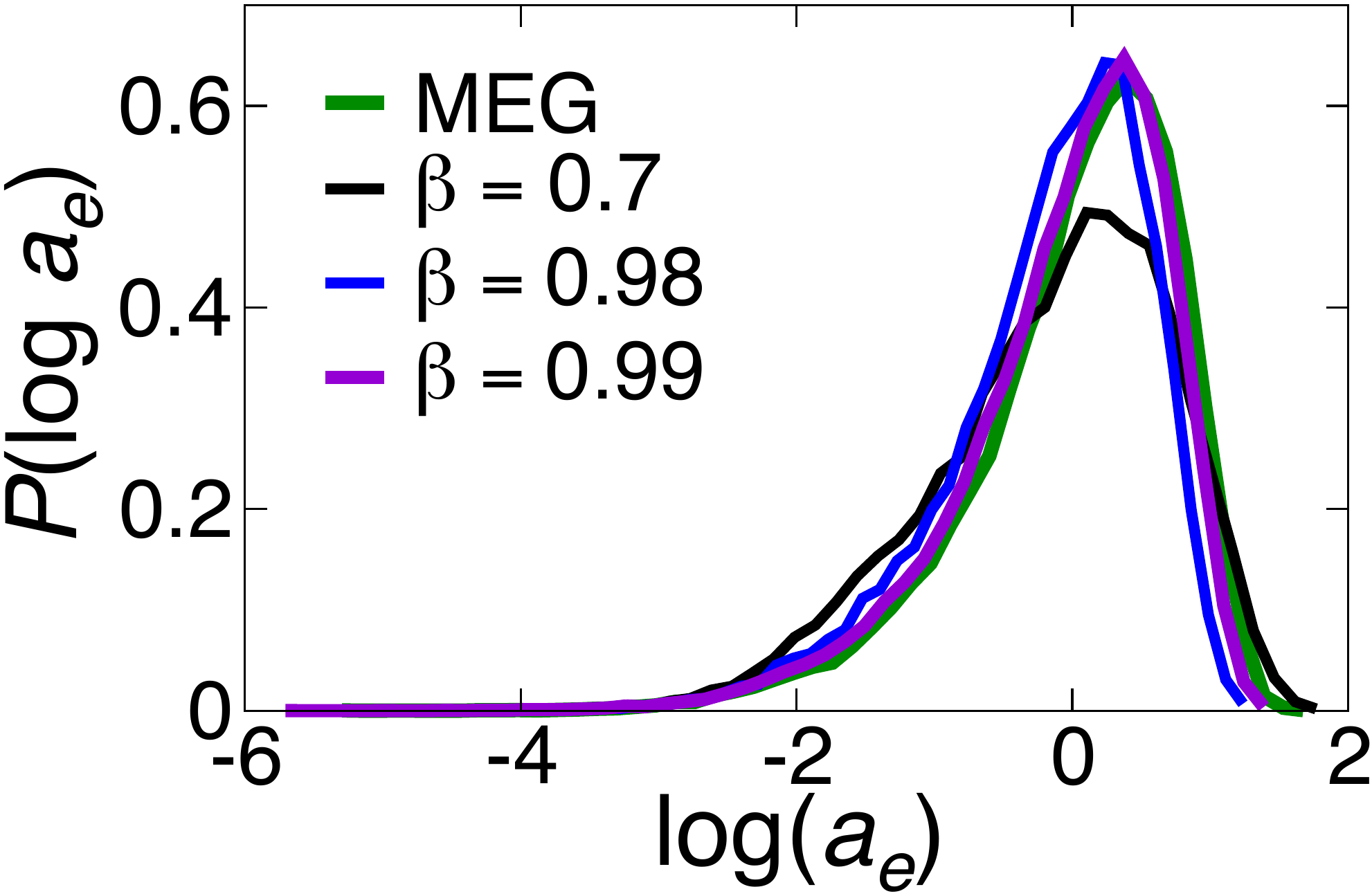}
\caption{{\bf Dependence of the distribution $P(\log a_e)$ on the value of $\beta$.} Distributions $P(\log a_e)$ of the logarithm of the area under the curve $a_e$, with $e = 2.5$SD, for MEG data (green curves = average over sensors for one subject) and the model with different $\beta$ values and $c = 0.5$.}
\label{SI_fig:6}
\end{figure}

\begin{figure}[ht!]
\centering
\includegraphics[width=\linewidth]{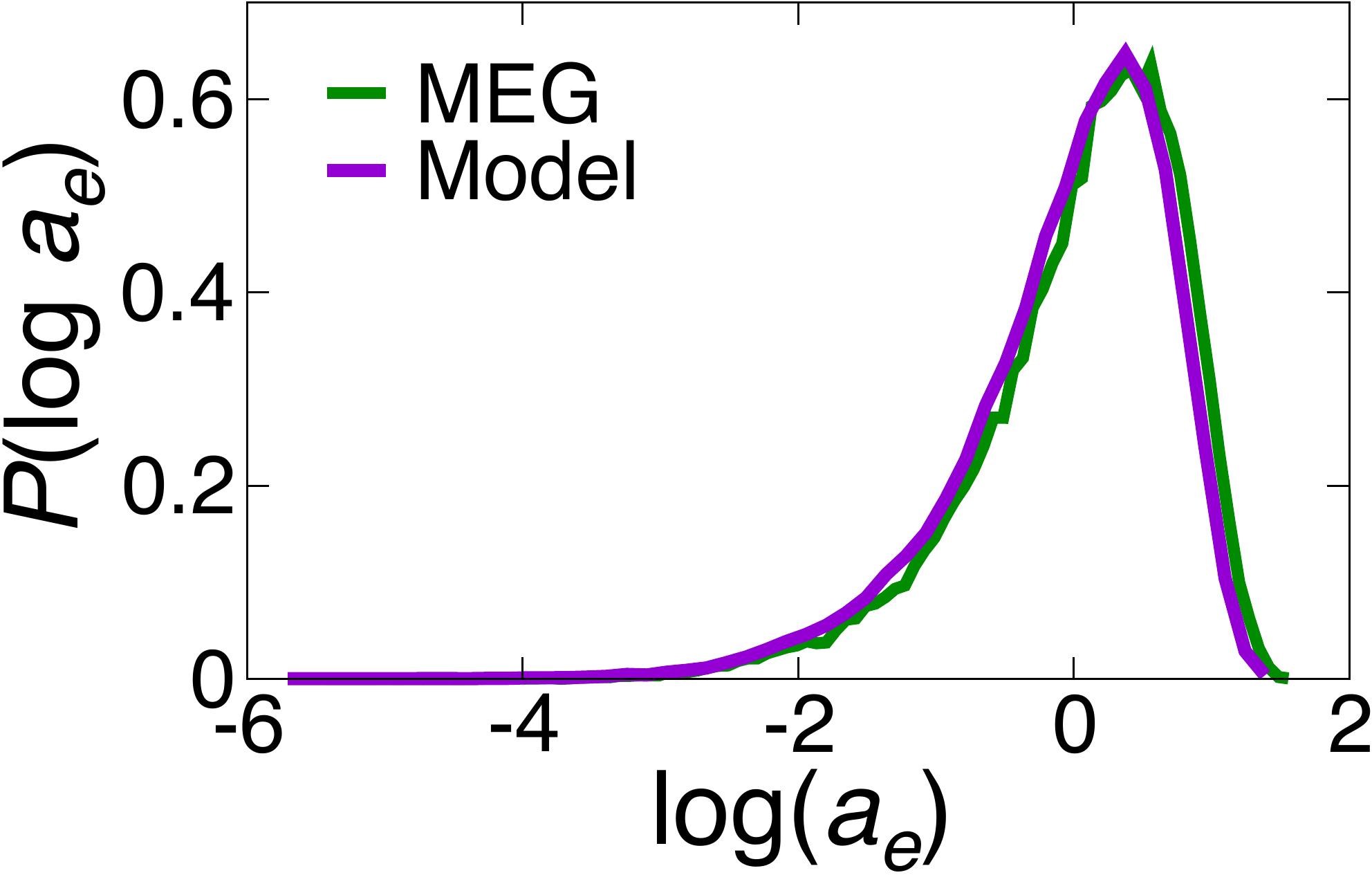}
\caption{{\bf Distributions $P(\log a_e)$ of the logarithm of the area under the curve $a_e$ for a single subject, with $e = 2.5$SD, for MEG data (green curves = average over sensors for one subject) and the model with corresponding inferred parameter values.}}
\label{SI_fig:7}
\end{figure}

\begin{figure}[ht!]
\centering
\includegraphics[width=\linewidth]{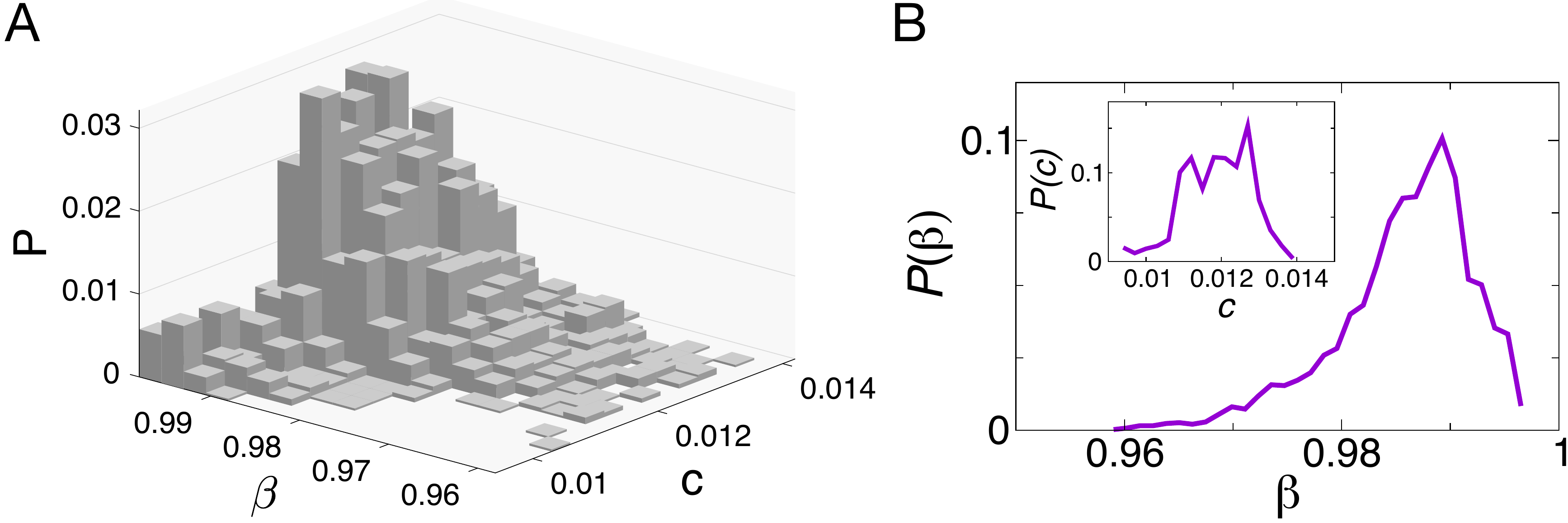}
\caption{{\bf Inferred parameter values.} (A) Joint probability of inferred values of the two model parameters from all MEG sensors (273) and subjects (14). (B) Marginal probability distributions of the  inferred model parameters $\beta$ (main panel) and $c$ (inset).}
\label{SI_fig:8}
\end{figure}

\begin{figure}[t!]
\centering
\includegraphics[width=\linewidth]{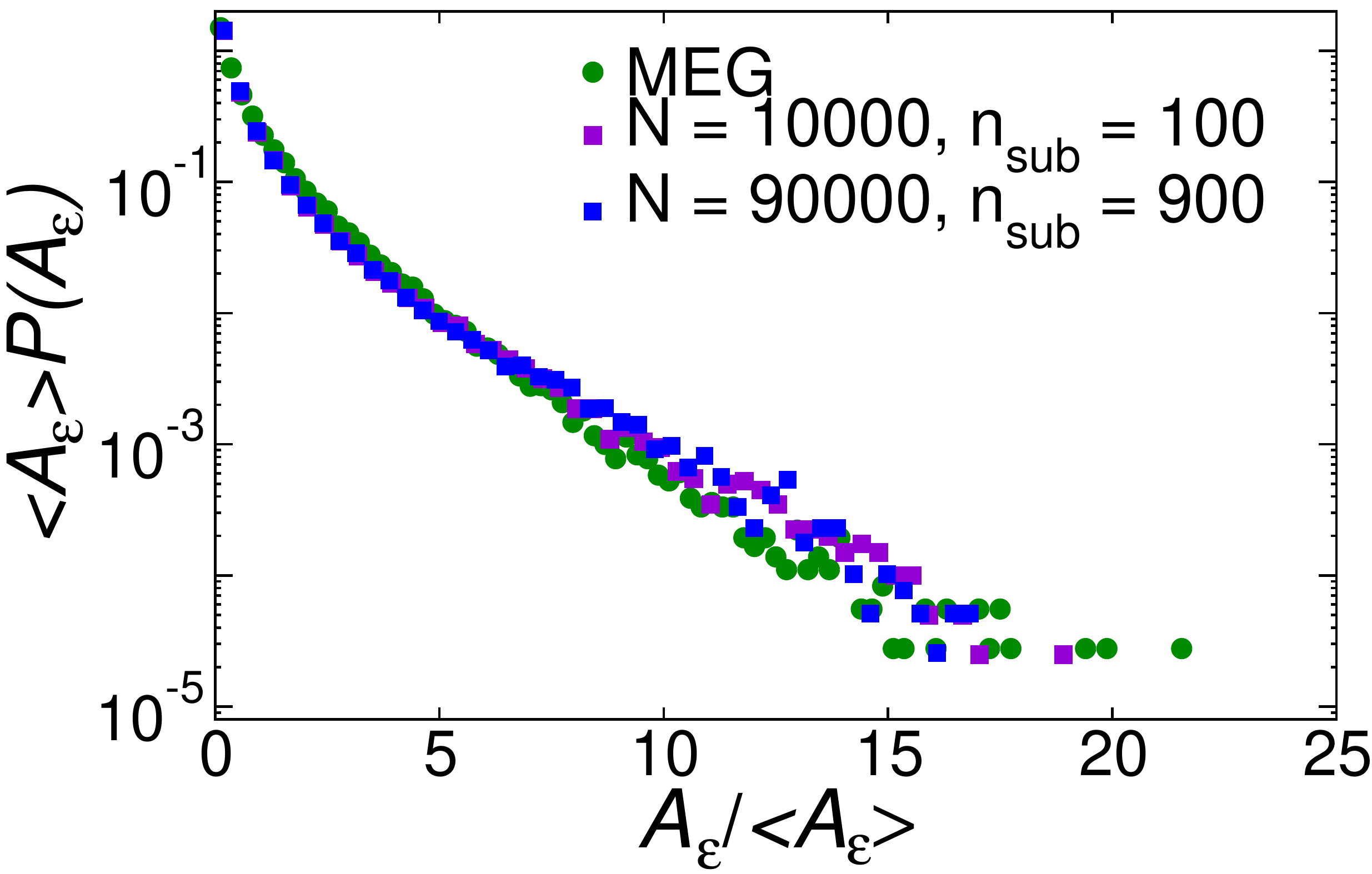}
\caption{{\bf Distributions of the activity per bin $A_{\epsilon}$   in model simulations with $N = 10^4$ and $N = 9\cdot 10^4$  spins ($\beta = 0.99$, $c = 0.01$).} The model network is parceled into subsystems of different  size $n_{\rm sub}$. The distribution $P(A_{\epsilon})$ is independent of the subsystem size $n_{\rm sub}$. The network excitation $A_{\epsilon}$ is rescaled by the average network excitation $\langle A_{\epsilon} \rangle$,  with $\epsilon = 2T$, where $T$ is the sampling interval as in Fig.~3.}
\label{SI_fig:9}
\end{figure}

\begin{figure}[t!]
\centering
\includegraphics[width=\linewidth]{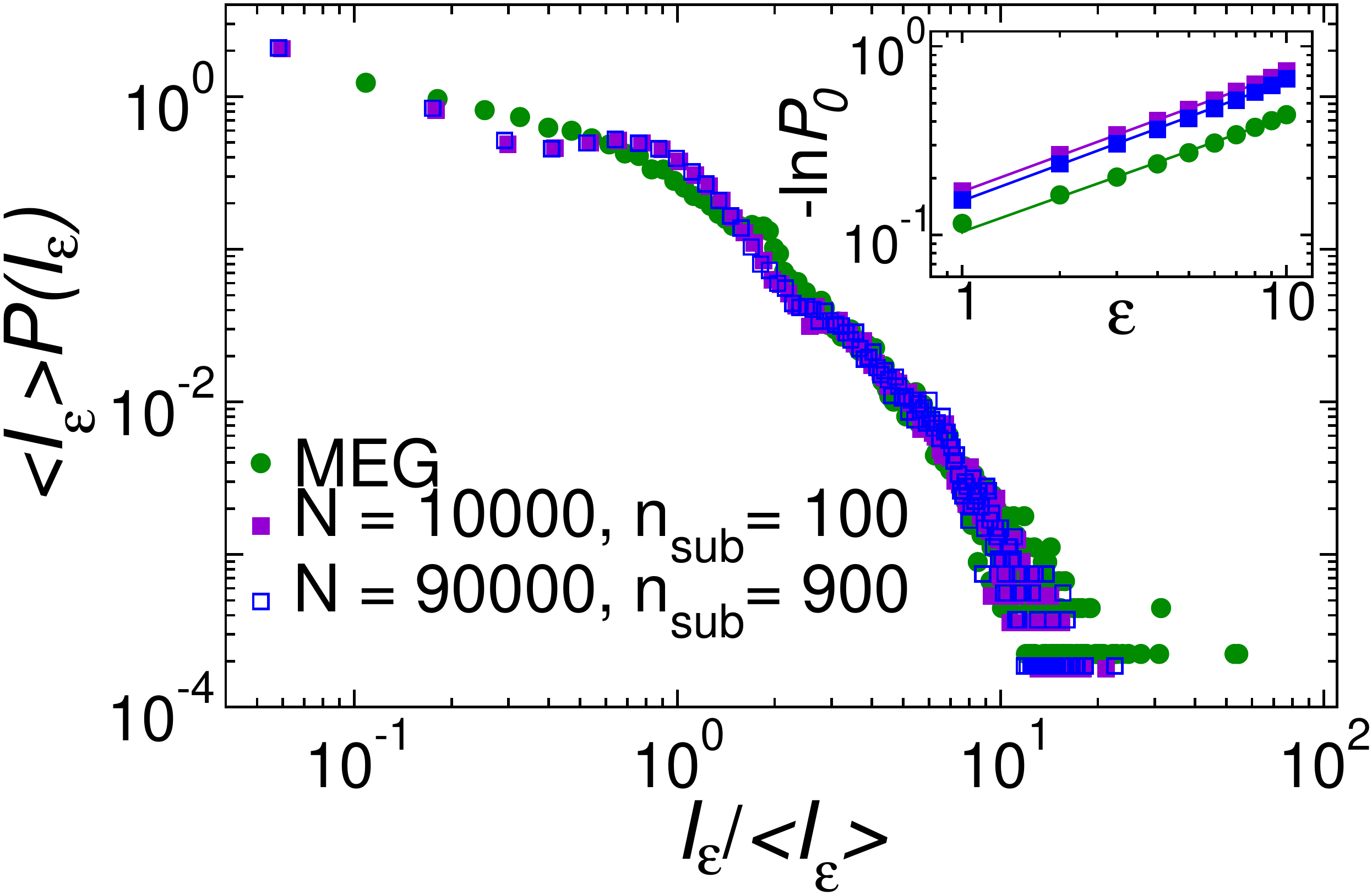}
\caption{{\bf Distributions of quiescence durations  $I_{\epsilon}$   in model simulations with $N = 10^4$ and $N = 9\cdot 10^4$  spins ($\beta = 0.99$, $c = 0.01$). } The model network is parceled into subsystems of different  size $n_{\rm sub}$. The distribution $P(I_{\epsilon})$ is independent of the subsystem size $n_{\rm sub}$. $\epsilon = 2T$, where $T$ is the sampling interval (see Fig.~3). The quiescence duration $I_{\epsilon}$ is rescaled by the average quiescence duration $\langle I_{\epsilon} \rangle$. Inset: Probability $P_0$ of finding a quiescent time bin scales approximately as $P_0 = \exp\left(-a\epsilon^{\beta_I}\right)$ with bin size $\epsilon$; $\beta_I \simeq 0.6$ independently of the subsystem size $n_{sub}$.}
\label{SI_fig:10}
\end{figure}

\begin{figure}[t!]
\centering
\includegraphics[width=\linewidth]{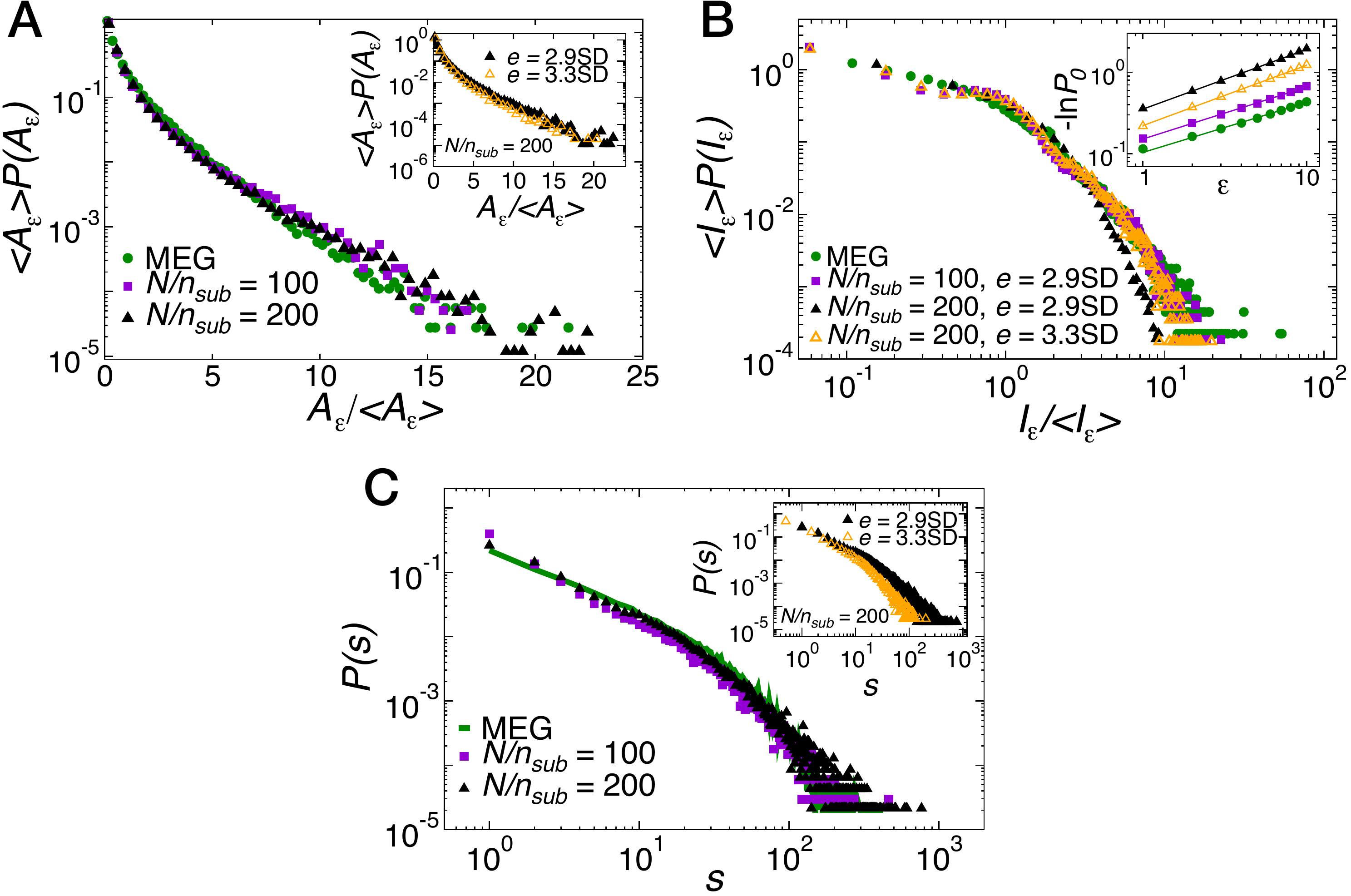}
\caption{{\bf Distribution $P(A_\epsilon)$ (A), $P(I_\epsilon)$ (B), and $P(s)$ (C) in model simulations with $N = 9 \cdot 10^4$  spins ($\beta = 0.99$, $c = 0.01$) and different numbers of subsystems $K = N/n_{sub}$}.  (A) The distribution of network excitation $P(A_\epsilon)$ ($\epsilon = \epsilon_2 = 2T$) weakly depends on the number of subsystems $K = N/n_{sub}$. Inset:  Distributions $P(A_\epsilon)$ for different values of the threshold $e$ used to detect extreme events. The number of subsystems is fixed to $N/n_{sub}= 200$.  (B) The distribution of  quiescence durations $P(I_\epsilon)$ ($\epsilon = \epsilon_2 = 2T$)  depends on the number of subsystems $K = N/n_{sub}$, particularly on the tail (black triangles up). For $N/n_{sub} = 200$ a good agreement between data and model simulations is recovered when the threshold $e$ is increased from $2.9$SD (the value used for $N/n_{sub} = 100$) to $3.3$SD (orange triangles up).  Inset: Probability $P_0$ of finding a quiescent time bin scales approximately as $P_0 = \exp\left(-a\epsilon^{\beta_I}\right)$ with bin size $\epsilon$; $\beta_I \simeq 0.6$ depends on the number of subsystems $N/n_{sub}$, and slightly increases from $\approx  0.6$ for $N/n_{sub}=100$ (violet squares) to $\approx 0.7$ for $N/n_{sub}=200$ (black triangles). (C) The distribution of avalanche sizes $P(s)$ weakly depends on the number $K = N/n_{sub}$. Inset:  Distributions $P(s)$ for different values of the threshold $e$ used to detect extreme events. The number of subsystems is fixed to $N/n_{sub}= 200$. Increasing $e$ has the effect of decreasing (increasing) the probability of large (small) avalanches. All $P(s)$ distributions are calculated with $\epsilon = \epsilon_4 = 4T$.}
\label{SI_fig:11}
\end{figure} 

\clearpage

\begin{figure}[t!]
\centering
\includegraphics[width=\linewidth]{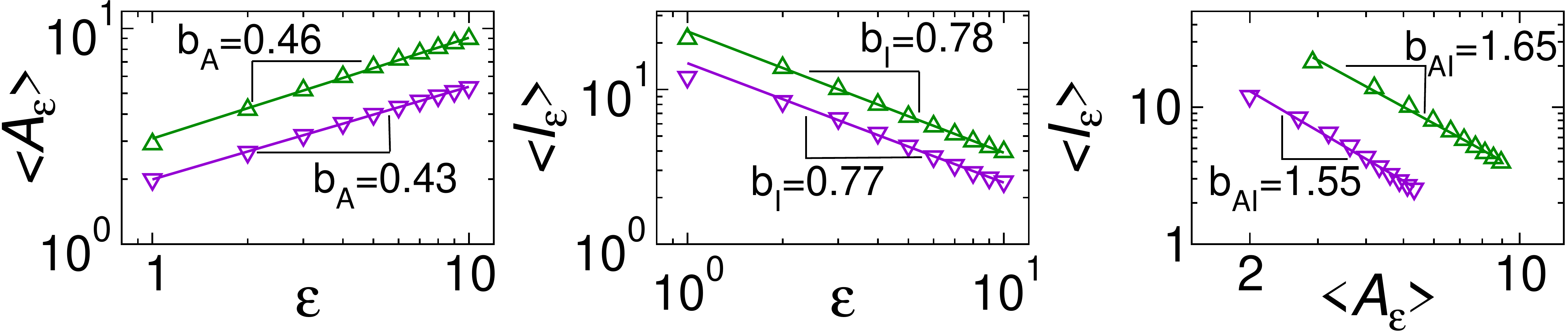}
\caption{{\bf The average network excitation, $\langle A_{\epsilon} \rangle$, and average quiescence duration,  $\langle I_{\epsilon} \rangle$, scale as a power-law of the bin size $\epsilon$, and are connected to each other by a power-law relationship. } (Left panel) $\langle A_{\epsilon} \rangle$ scales with $\epsilon$ as $\langle A_{\epsilon} \rangle \sim \epsilon^{b_A}$, with similar exponents $b_A$ in data and model simulations (Data: $b_A = 0.46 \pm 0.02$; Model: $b_A = 0.43 \pm 0.01$). (Middle panel) $\langle I_{\epsilon} \rangle$ scales with $\epsilon$ as $\langle I_{\epsilon} \rangle \sim \epsilon^{b_I}$, with similar exponents $b_I$ in data and model simulations (Data: $b_I = 0.78 \pm 0.03$; Model: $b_A = 0.77 \pm 0.01$). (Right panel)  $\langle I_{\epsilon} \rangle$ is connected to  $\langle A_{\epsilon} \rangle$ by the relationship $\langle I_{\epsilon} \rangle \sim \langle A_{\epsilon} \rangle^{b_{AI}}$, with similar exponents $b_{AI}$ in data and model simulations (Data: $b_{AI} = 1.65 \pm 0.12$; Model: $b_{AI} = 1.55 \pm 0.03$). Simulations are from a model with $N=90000$ and $K = 100$. Extreme events are extracted using a threshold $e = 2.9$SD in both data and model simulations. For a given $\epsilon$, the value of  $\langle A_{\epsilon} \rangle$ and $\langle I_{\epsilon} \rangle$ can be controlled adjusting the threshold $e$. This implies that, beside the scaling exponents, it is possible to match the values of those quantities in the model by appropriately tuning $e$.}
\label{SI_fig:12}
\end{figure} 

\clearpage

\begin{figure}[ht!]
\centering
\includegraphics[width=\linewidth]{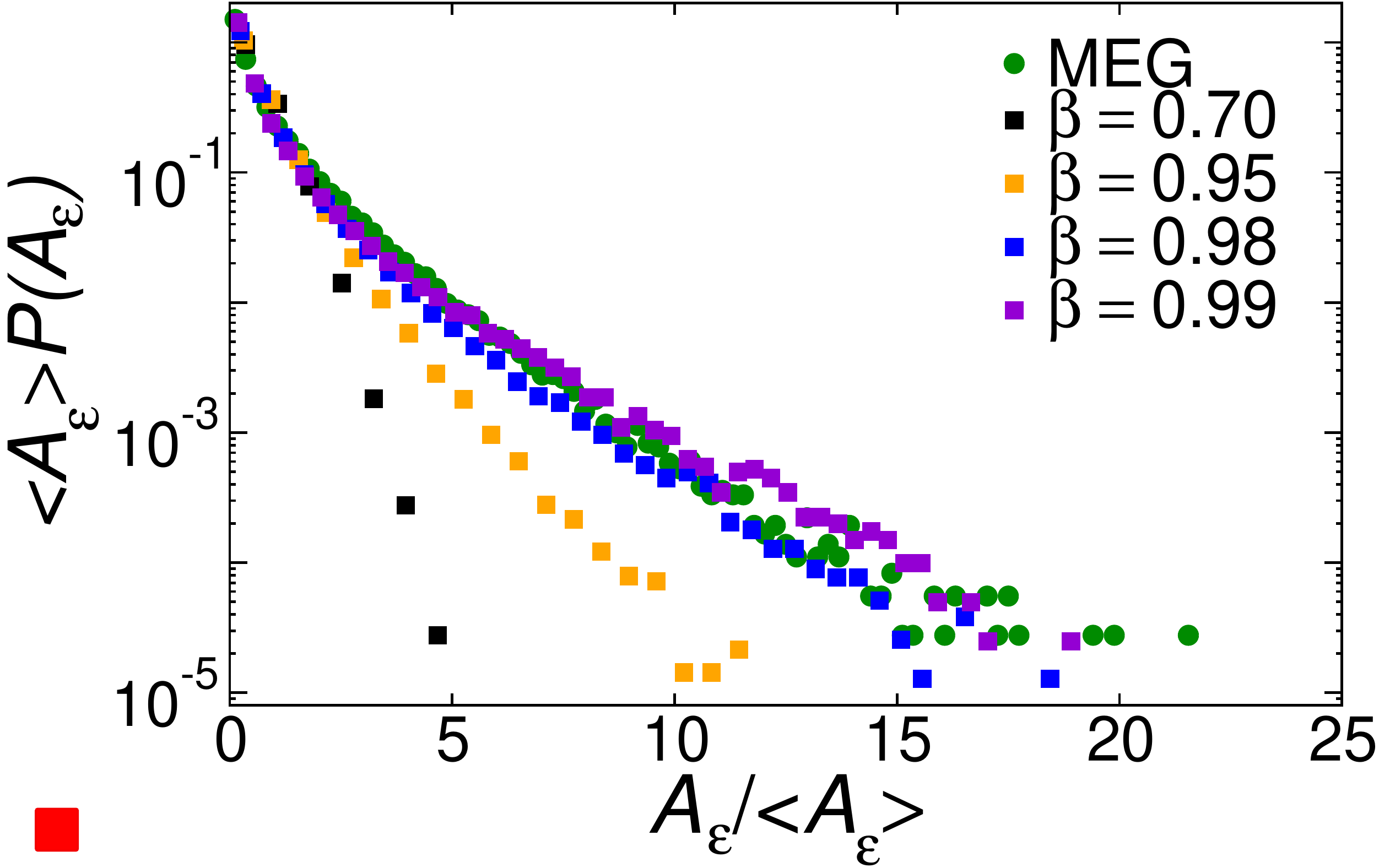}
\caption{{\bf Distributions of the activity per bin $A_{\epsilon}$ for different values of  $\beta$ in model simulations with $N=10^4$ spins, and for MEG data (average over subjects).} The model network is parceled in 100 disjoint subsystems, each including 100 spins. In all cases the model is in the resonant regime. The network excitation  $A_{\epsilon}$ is rescaled by the average network excitation $\langle A_{\epsilon} \rangle$ ($\epsilon = \epsilon_2 = 2T$, where $T$ is the sampling interval, as in Fig.~3). $\beta = 0.99$ corresponds to the average $\beta$ value inferred from MEG data.}
\label{SI_fig:13}
\end{figure}

\begin{figure}[ht!]
\centering
\includegraphics[width=\linewidth]{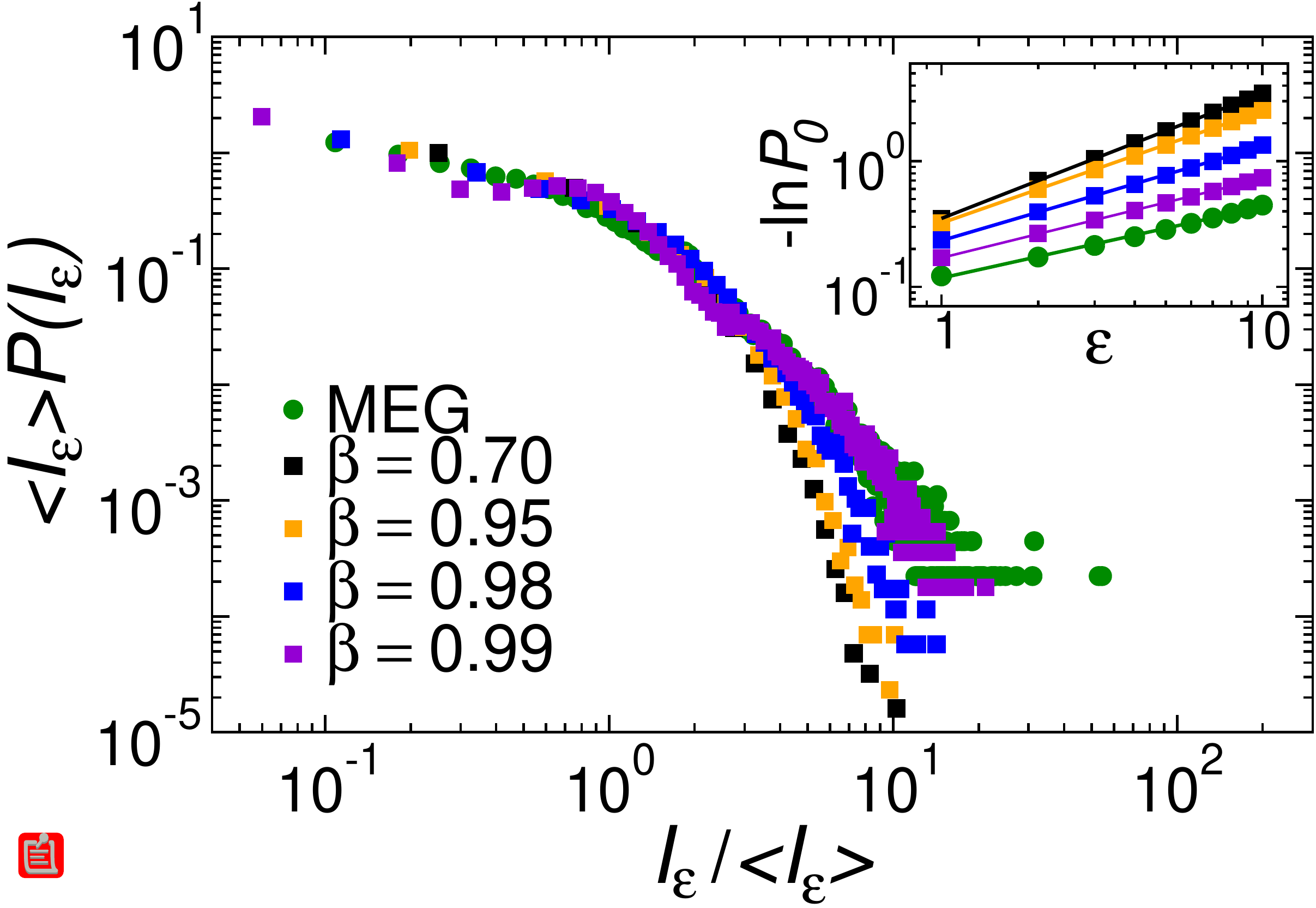}
\caption{{\bf Distributions of quiescence durations  $I_{\epsilon}$ for different values of  $\beta$ in model simulations with $N=10^4$ spins, and for MEG data (average over subjects).} The model network is parceled in 100 disjoint subsystems, each including 100 spins. In all cases the model is in the resonant regime.  The quiescence duration $I_{\epsilon}$ is rescaled by the average quiescence duration $\langle I_{\epsilon} \rangle$ ($\epsilon = \epsilon_2 = 2T$, where $T$ is the sampling interval, as in Fig.~3). $\beta = 0.99$ corresponds to the average $\beta$ value inferred from MEG data. Inset: Probability $P_0$ of quiescence periods as a function of $\epsilon$ for different $\beta$ values. We notice that as we move away from the critical point $\beta_c = 1$, the probability tends to follow an exponential behavior, i.e. $P_0 \propto e^{-a \epsilon}$. On the other hand, for $\beta = 0.99$ we find $P_0 \propto e^{-a \epsilon ^{\beta _I}}$, with $\beta _I = 0.6304 \pm 0.0046$, close to the value measured in MEG data (green circles) ($\beta _I = 0.5669 \pm 0.0117$)).}
\label{SI_fig:14}
\end{figure}

\begin{figure}[ht!]
\centering
\includegraphics[width=\linewidth]{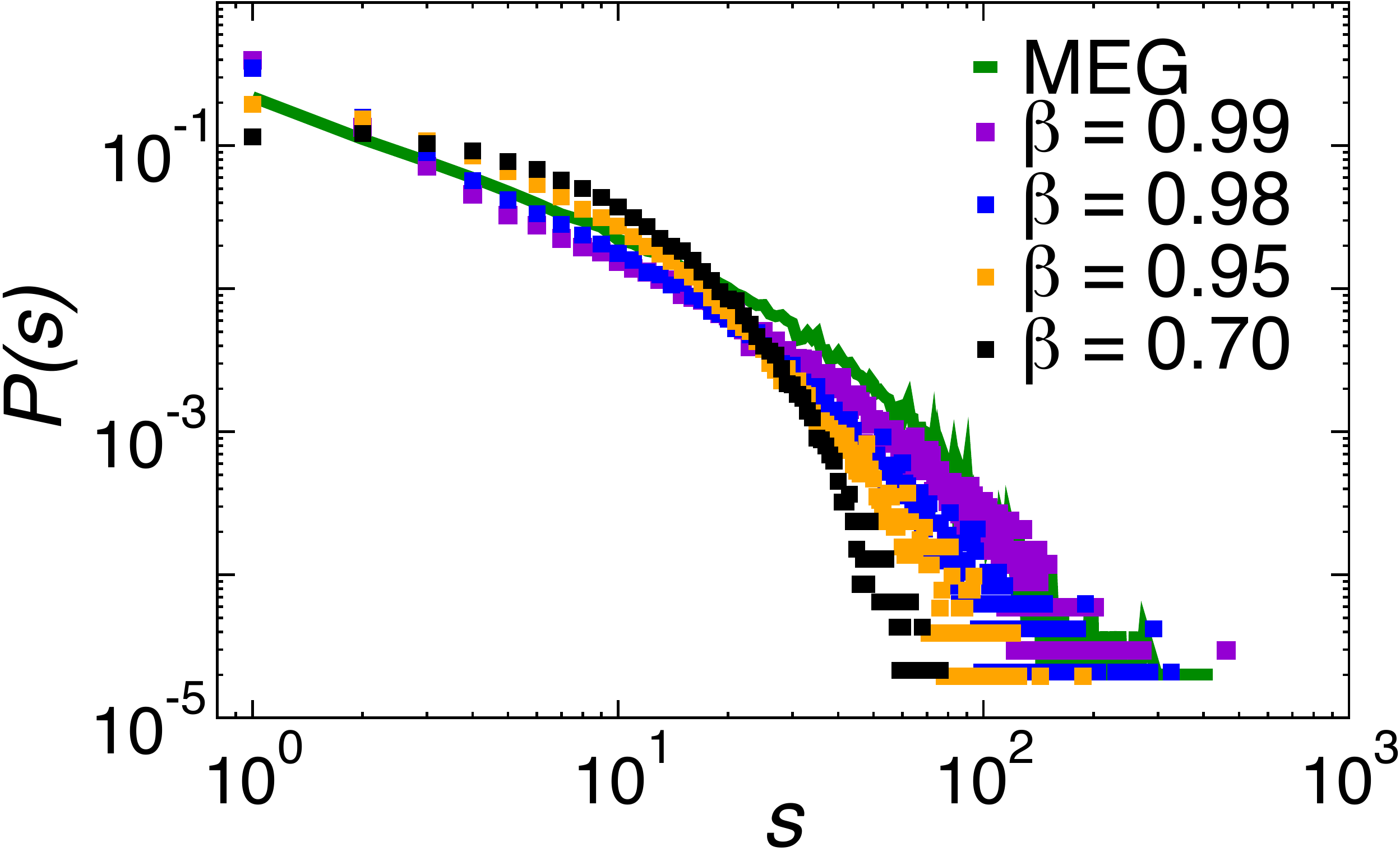}
\caption{{\bf Distribution of avalanche sizes, $P(s)$, for MEG data (green curve = average over subjects) and the model simulated at different $\beta$ values in the resonant regime, i.e. $c > c^*$.} Distributions are estimated using a threshold $e = 2.9$SD and bin size $\epsilon_4 = 4T$. Already for $\beta = 0.95$, a value slightly smaller than the baseline value $0.99$, avalanche sizes from model simulations tend to follow an exponential distribution that is far from reproducing avalanche size distributions from MEG data.}
\label{SI_fig:15}
\end{figure}

\begin{figure}[t!]
\centering
\includegraphics[width=\linewidth]{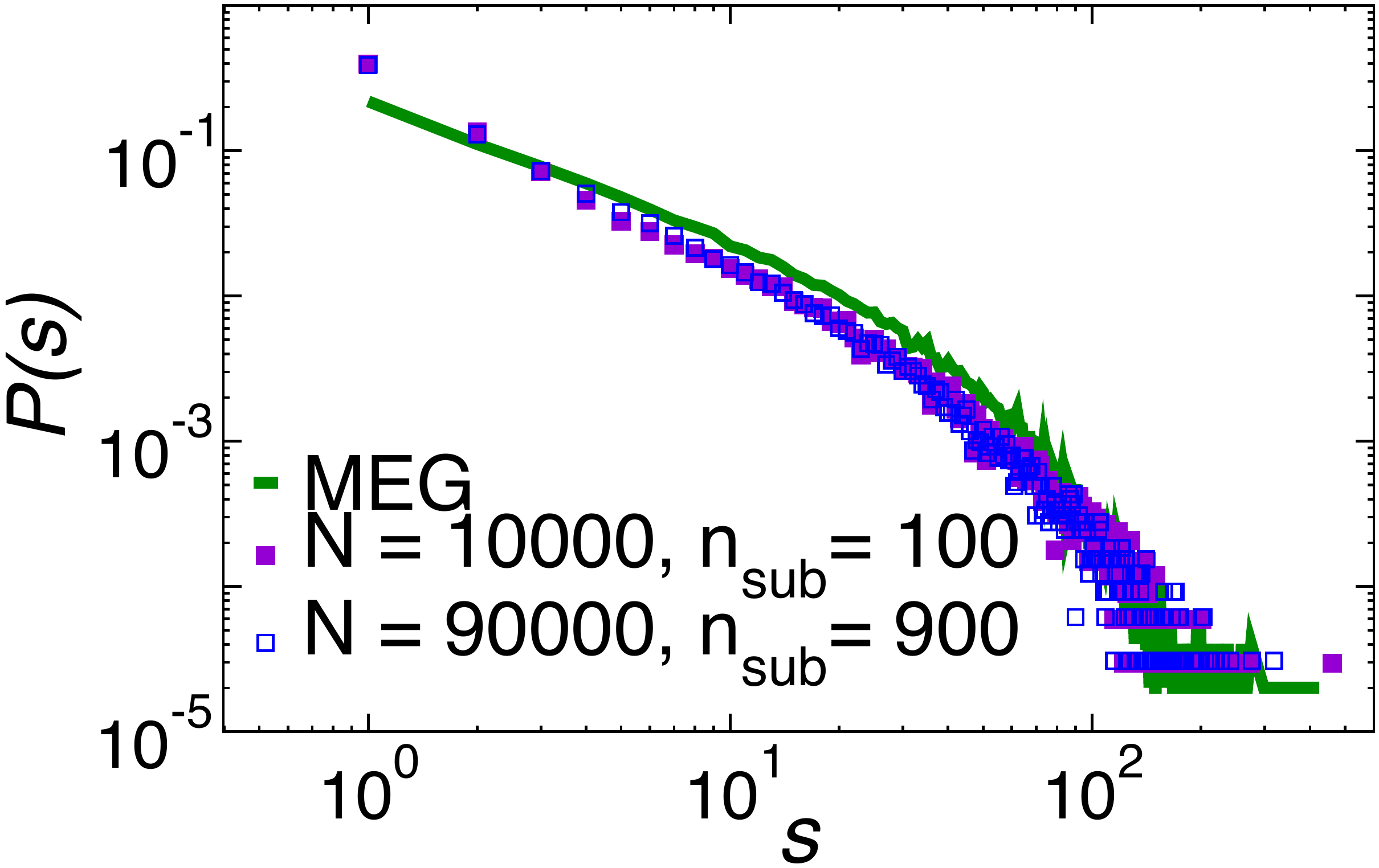}
\caption{{\bf Distribution $P(s)$ of avalanche sizes   in model simulations with $N = 10^4$ and $N = 9\cdot 10^4$ spins ($\beta = 0.99$, $c = 0.01$).} The model network is parceled in subsystems of different  size $n_{\rm sub}$. The distribution $P(s)$ is independent of the subsystem size $n_{\rm sub}$.}
\label{SI_fig:16}
\end{figure}

\end{document}